\documentclass{vldb}
\usepackage{graphicx,verbatim,fancyvrb,listings,
subfigure,algorithmic,algorithm,enumerate,stfloats,mdframed}
\usepackage{balance}  % for  \balance command ON LAST PAGE  (only there!)
\begin{document}
%% Prelim chapter

%\lstset{numbers=left}
\lstdefinelanguage{AQL}{
  alsoletter={-"'_:\$[]()0123456789\\},
  morekeywords={
    :,:=    
    for,
    let,
    in,
    return,
    where,
    group,
    by,
    with,
    order,
    desc,
    asc,
    limit,
    replace,
    delete,
    insert,
    some,
    every,
    satisfies,
    and,
    or,
    to,
    set,
    simfunction,
    simthreshold,
    secondary,
    feed,
    disconnect,
    where,
    some,
    satisfies,
    group,
    by,
    return,
  },
  basicstyle=\sf,
  keywordstyle=\textbf,
  identifierstyle=\texttt,
  commentstyle=\textit,
  literate={<=}{{\litleq}}1 {>=}{{\litgeq}}1,
  literate={~=}{$\sim{}=$}1 % set tilde as a literal (no process)
}[keywords,comments,strings]

\lstdefinelanguage{AQLSchema}{
  alsoletter={-"'_:(\$\\},
  morekeywords={
    :,:=,=,``,'',(,),    
    for,let,in,    
    use,
    declare,
    type,
    as,
    open,
    closed,
    dataset,
    dataverse,
    nodegroup,
    on,
    primary,
    string,double,int32, datetime, timestamp,point,
    ?,
    key,
    create,
    index,
    using,
    apply,
    function,
    feed,
    connect,
    to,
    from,
    policy,
    secondary,
    disconnect,
    where,
    some,
    satisfies,
    group,
    by,
    with,
    return,
    rtree,
    set,
  },
  basicstyle=\sf,
  keywordstyle=\textbf,
  identifierstyle=\sf,
  commentstyle=\textit,
  literate={<=}{{\litleq}}1 {>=}{{\litgeq}}1
}[keywords,comments,strings]

\lstnewenvironment{aql}
   {\lstset{language=AQL,columns=space-flexible,basicstyle=\small,basewidth={.5em,.5em}}}
   {}

\lstnewenvironment{aqlscriptsize}
   {\lstset{language=AQL,columns=space-flexible,basicstyle=\scriptsize,basewidth={.5em,.5em},numbers=left}}
   {}

\lstnewenvironment{aqltiny}
   {\lstset{language=AQL,columns=space-flexible,basicstyle=\tiny,basewidth={.5em,.5em},numbers=left}}
   {}

\lstnewenvironment{aqlschema}
   {\lstset{language=AQLSchema,columns=space-flexible,basicstyle=\small,basewidth={.5em,.5em}}}
   {}

% THIS IS AN EXAMPLE DOCUMENT FOR VLDB 2012
% based on ACM SIGPROC-SP.TEX VERSION 2.7
% Modified by  Gerald Weber <gerald@cs.auckland.ac.nz>
% Removed the requirement to include *bbl file in here. (AhmetSacan, Sep2012)
% Fixed the equation on page 3 to prevent line overflow. (AhmetSacan, Sep2012)

% ****************** TITLE ****************************************

\title{Scalable Fault-Tolerant Data Feeds in AsterixDB %\\[.5ex] \textsf{\Large[Innovative Systems Paper]}%
} 
\vspace{-20mm}
\author{
% 1st. author
\alignauthor Raman Grover  {\small $~^{1}$}, Michael J. Carey {\small $~^{2}$} \\
       \fontsize{12}{12}
       \email{\emph{\{ramang$~^{1}$, mjcarey$~^{2}$\}@ics.uci.edu}}\\
	   \vspace{1.7mm}     
       \affaddr{Department of Computer Science,}\\
       \vspace{0.5mm}
       \affaddr{University of California, Irvine CA 92697 USA}\\
}
\maketitle

\begin{abstract} 
In this paper we describe the support for data feed ingestion in AsterixDB, an open-source Big Data Management System (BDMS) that provides a platform for storage and analysis of large volumes of semi-structured data. Data feeds are a mechanism for having continuous data arrive into a BDMS from external sources and incrementally populate a persisted dataset and associated indexes. The need to persist and index ``fast-flowing'' high-velocity data (and support ad hoc analytical queries) is ubiquitous. However, the state of the art today involves `gluing' together different systems. AsterixDB is different in being a unified system with ``native support'' for data feed ingestion. 

We discuss the challenges and present the design and implementation of the concepts involved in modeling and managing data feeds in AsterixDB. AsterixDB allows the runtime behavior, allocation of resources and the offered degree of robustness to be customized to suit the high-level application(s) that wish to consume the ingested data. Initial experiments that evaluate scalability and fault-tolerance of AsterixDB data feeds facility are reported. 
\end{abstract}

\section{Introduction}

A large volume of data is being generated on a ``continuous'' basis, be it in the form of click-streams, output from sensors, log files or via sharing on popular social websites \cite{ BigDataRise}. Encouraged by low storage costs, data-driven enterprises today are aiming to collect and persist the available data and analyze it over time to extract hidden insightful information. Marketing departments use Twitter feeds to conduct sentiment analysis to determine what users are saying about the company's products. 
\begin{comment}
Location data combined with customer preference data from social networks enable retailers to target marketing campaigns based on buying history. 
\end{comment}
As another example, utility companies have rolled out smart meters that measure the consumption of water, gas, and electricity and generate huge volumes of interval data that is required to be analyzed over time.

Traditional data management systems require data to be loaded and indexes be created before data can be subjected to ad hoc analytical queries. To keep pace with ``fast-moving'' high-velocity data, a Big Data Management System (BDMS) must be able to ingest and persist data on a continuous basis. A flow of data from an external source into persistent (indexed) storage inside a BDMS will be referred to here as a \emph {data feed}. The task of maintaining the continuous flow of data is hereafter referred to as \emph{data feed management}. 

A simple way of having data being put into a Big Data management system on a continuous basis is to have a single program (process) fetch data from an external data source, parse the data and then invoke an insert 
statement per record/batch. This solution is limited to a single machine's fetching/computing capacity. Ingesting multiple data feeds would potentially require running and managing individual programs/processes.
The task of continuously retrieving data from external source(s), applying some pre-processing for cleansing, filtering or transforming data may amount to `gluing' together different systems (e.g. \cite{HadoopDW}). It becomes hard to reason about data consistency, scalability and fault-tolerance offered by such an assembly. Traditional data management systems have evolved to provide native support for services if the service offered by an external system is inappropriate or may cause substantial overheads \cite{Stonebraker:1981:OSS:358699.358703, SensorDatabases}. Responding to the need of the hour then, it is natural and efficient for a BDMS to provide ``native'' support for data feed management. 
%\vspace{-1mm}
\subsection{Challenges in Data Feed Management} 
\label{subsec:challenges}
Let us begin by enumerating the challenges involved in building a data feed ingestion facility and emphasize on the desirable features of such a system.  

C1)  \emph{Genericity} and \emph{Extensibility}:
A feed ingestion facility must be generic enough to work with a variety of data sources and high-level applications. A plug-n-play model is desired to allow modification of the offered functionality.  

C2)  \emph{Fetch-Once Compute-Many Model}:
A data feed could simultaneously drive multiple applications that might require the arriving data to be processed/persisted differently. It is desirable to have a single flow of data from an external source and yet be able to transform it in multiple ways to drive different applications concurrently. 

C3) \emph{Data Feed Monitoring and Resource Management}: 
Multiple feeds may be concurrently active, each competing for resources to keep pace with its data source. A concurrent query may further increase the demand for resources. It is essential to monitor each feed and attempt to prevent/resolve bottlenecks by efficient allocation of resources.

C4) \emph{Fault Tolerance}:
Data feed ingestion is expected to run over commodity hardware and is therefore prone to hardware failures. Unexpected format or value(s) for attributes in data may cause soft failures. The feed ingestion facility should offer the desired degree of robustness in handling failures while minimizing data loss. 

C5) \emph{Scalability}:
The system should be able to ingest an increasingly large volume of data, possibly from multiple data feeds in parallel when additional resources are added.  

\subsection{Contributions}
In this paper, we describe the support for data feed management in AsterixDB and discuss the approach adopted to address the  challenges from Section~\ref{subsec:challenges}.
AsterixDB is a Big Data Management System (BDMS) that provides a platform for the scalable storage and analysis of very large volumes of semi-structured data. The paper offers the following contributions. 

%\vspace{1mm}
(1) \emph{Concepts involved in Data Feed Management}:
The paper introduces the concepts involved in modeling and defining a data feed and managing the flow of data into a target dataset and/or to other dependent feeds to form a cascade network.
It includes the design and implementation of the involved concepts in a complete system. 

(2) \emph{Policies for Data Feed Management}:
We describe how a data feed is managed by associating an ingestion policy that controls the runtime behavior in response to events such as software/hardware failures and resource bottlenecks. Users may also provide a custom policy to suit special application requirements. 

(3) \emph{Fault-Tolerant Data Feed Management}: 
We provide a taxonomy of failures and provide a fault-tolerance protocol that can be customized as per the application requirements to provide the desired degree of robustness. 

(4) \emph{Contribution to Open-Source}:
AsterixDB is available as open source \cite{AsterixDBSource, AsterixDB}. The support for data feed ingestion in AsterixDB is extensible so that future contributors can provide custom implementation of different modules and form custom-designed policies to suit specific requirements.

(5) \emph{Experimental Evaluation}: 
We provide an initial experimental evaluation and study the ability of the system to scale and ingest increasingly large volume of data with the addition of resources. We also report experiments to evaluate our fault-tolerance approach under different failure scenarios.  

The remainder of the paper is organized as follows. We discuss related work in Section \ref{sec:related} and provide an overview of the AsterixDB system in Section \ref{sec:background_asterixdb_hyracks}. Section \ref{sec:modeling_a_data_feed} describes how a feed is modeled and defined at a language level in AsterixDB. 
The implementation details involved in managing a data feed are described in Section~\ref{sec:runtime_for_feed_ingestion}. Section \ref{sec:fault_tolerant_feed_ingestion} describes the support for handling failures. Section \ref{sec:experimental_evaluation} provides an experimental evaluation. Finally, Section \ref{sec:conclusion} concludes the paper and discusses future work.
%\vspace{-0.2cm}

\section{Related Work}
\label{sec:related}
Data feeds may seem similar to streams from the data streams literature (e.g. \cite{Abadi:2003:ADS:872757.872855, Spade}). However, there are important differences. Data feeds are a ``plumbing'' concept; they are a mechanism for having data flow from external sources that produce data continuously and to incrementally populate and persist in a database. Stream Processing Engines (SPEs) do not persist data; instead they provide a sliding window on data (e.g. a 2 minute or 5 minute view of data), but the amount, or the time window is usually limited by the velocity of the data and the available memory. 
In a similar spirit, Complex Event Processing (CEP) systems (Storm \cite{Storm}, S4 \cite{S4}) can route, transform and analyze a stream of data. However these systems do not persist the data or provide support for ad hoc analytical queries. These engines can be used in conjunction with a database (e.g MySql), making it possible to persist and run ad hoc queries but with the assumption that the target database can handle high-velocity data. 

In the past, ETL (Extract Transform Load) systems (e.g. \cite{INFORMATICA_ETL}) have dealt with the challenge of populating a Data Warehouse with data collected from multiple data sources. However, such systems operate in a ``batchy'' mode, with a ``finite'' amount of data transferred at periodic intervals coinciding with off-peak hours. Xu in \cite{HadoopDW} described a Map-Reduce based approach for populating a parallel database system with data arrving from an external source. However, the system formed a tight coupling with Map-Reduce and required data to be initially put into HDFS. 

With respect to providing fault-tolerance, stream processing was also met with the challenge of providing highly available parallel data-flows and have proposed several techniques \cite{Shah:2004, BorealisFT}. These techniques rely on replication where the state of an operator is replicated on multiple servers or have multiple servers simultaneously process the identical input streams. Fault-tolerance is provided at a high cost as the number of nodes are at least reduced to half. Moreover, offering a single strategy for fault-tolerance can be wasteful of resources in scenarios where the offered degree of robustness exceeds the requirements. 

The need to be able to persist and index fast-flowing data is ubiquitous. Instead of `gluing' together different systems, AsterixDB is different in being a unified system with ``native'' support for data feed ingestion. AsterixDB offers a generalized fault-tolerance approach that can be customized as per the application requirements and the expected degree of robustness. Furthermore, AsterixDB does not require data to be staged to external storage (e.g. HDFS) before being ingested. To the best of our knowledge, AsterixDB is the first system to explore the challenges involved in building a data ingestion facility that is fault tolerant and employs partitioned parallelism to scale the facility and couple it with high-volume and/or parallel external data sources. The most closely related work is the AT\&T Bistro data feed management system \cite{Shkapenyuk:2011:BDF:1989323.1989437}, but that work focused on routing large amounts of file-based data from pre-determined feeds to the applications that need access to them. 
\vspace{-1mm}

\section{Background: AsterixDB}
\label{sec:background_asterixdb_hyracks}

Initiated in 2009, the NSF-sponsored ASTERIX project has been developing new technologies for ingesting, storing, indexing, querying, and analyzing vast quantities of semi-structured data. The project has combined ideas from three distinct areas---semi-structured data, parallel databases, and data-intensive computing---in order to create an open-source software platform that scales by running on large, shared-nothing commodity computing clusters. 

\subsection{AsterixDB Architecture}
\begin{figure}[!h]
  \includegraphics[width=90mm,height=45mm]{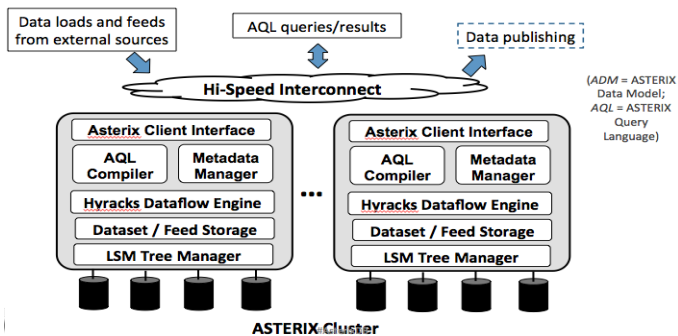}
  \caption{AsterixDB Architecture}
  \label{fig:AsterixArchitecture}
\end{figure}
Figure ~\ref{fig:AsterixArchitecture} provides an overview of how the various software components of AsterixDB map to nodes in a shared-nothing cluster. The topmost layer of AsterixDB is a parallel DBMS, with a full, flexible data model (ADM) and query language (AQL) for describing, querying, and analyzing data. ADM and AQL support both native storage and indexing of data as well as analysis of external data (e.g., in HDFS). The bottom-most layers from Figure~\ref{fig:AsterixArchitecture} provide storage facilities for datasets, which can be targets of ingestion. These datasets are stored and managed by AsterixDB as partitioned LSM-based B+-trees with optional LSM-based secondary indexes. 

AsterixDB uses Hyracks \cite{DBLP:conf/icde/BorkarCGOV11} as its runtime execution layer. Hyracks allows AsterixDB to express a computation as a DAG of data operators and connectors. Operators operate on partitions of input data and produce partitions of output data. Connectors repartition operators' outputs to make the newly produced partitions available at the consuming operators. Hyracks sits at roughly the same level that Hadoop (Map-Reduce) does in implementations of other high-level languages such as Pig, Hive or Jaql. 

\subsection{AsterixDB Data Model} 
\label{subsec:AsterixDB_Data_Model}
AsterixDB defines its own data model (ADM) \cite{AsterixVision} that was designed to support rich semi-structured data with support for bags/lists, nested types and a variety of primitive types. Figure~\ref{fig:feed_data_specification} illustrates ADM by showing how it could be used to define a record type for modeling a raw tweet\footnote{At the time of writing, this best reflects a real tweet obtained from the Twitter API.}. The record type shown there is an open type, meaning that its instances will conform to its specification but can contain extra fields that vary from instance to instance.  Figure~\ref{fig:feed_data_specification} also includes the definition of a processed tweet. A processed tweet replaces the nested user field inside a raw tweet with a primitive userId value and adds a nested collection of strings (referred topics) to each tweet. The primitive field types (location-lat, location-long) and send-time are expressed as their respective spatial  (point) and temporal (datetime) datatypes. ADM also allows specifying optional fields with known types (e.g. sender-location). 
\begin{figure}
\floatstyle{boxed}
\vspace{-0.6cm}
\begin{aqlschema}

create type RawTweet         create type TwitterUser 
as open {                    as open { 
   tweetId: string,             screen-name: string, 
   user: TwitterUser,           lang: string,
   location-lat: double?,       friends_count: int32,
   location-long: double?,      statuses_count: int32,
   send-time: string,           name: string, 
   message-text: string         followers_count: int32
};                           };  
                                   
create type ProcessedTweet as open {
   tweetId: string,
   userId: string,
   sender-location: point?,
   send-time: datetime,
   message-text: string,
   referred-topics: {{string}}
};

\end{aqlschema}
\vspace{-0.6cm}
\caption{Defining datatypes}
\label{fig:feed_data_specification}
\vspace{-0.4cm}
\end{figure}

\begin{figure}[htb]
%\vspace{-0.3cm}
\begin{aqlschema}

create dataset RawTweets(RawTweet)
primary key tweetId;

create dataset ProcessedTweets(ProcessedTweet)
primary key tweetId;

create index locationIndex on 
ProcessedTweets(location) type rtree;

\end{aqlschema}

\vspace{-0.2cm}
\caption{Creating datasets and associated indexes}
\label{fig:feed_dataset_definition}
\end{figure}
%\vspace{-0.4cm}

Data in AsterixDB is stored in \emph{datasets}. Each record in a dataset conforms to the datatype associated with the dataset. Data is hash-partitioned (primary key) across a set of nodes that form the \emph{nodegroup} for a dataset\footnote{By default, the nodegroup for a dataset includes all nodes in an AsterixDB cluster.}.  Figure~\ref{fig:feed_dataset_definition} shows the AQL statements for creating a pair of datasets---\emph{RawTweets} and \emph{ProcessedTweets}. 
Additionally we create a secondary index on the location attribute of a processed tweet for efficient retrieval/grouping of tweets on the basis of spatial location.

\subsection{Querying Data} 
AsterixDB queries are written in AQL, a declarative query language that was designed by taking the essence of XQuery \cite{XQuery}. As an example, consider the AQL query in Figure~\ref{fig:AQL_example_query} which spatially aggregates tweets collected in the dataset \emph{ProcessedTweets}. The query defines a bounding rectangle that spans over the geographic region covered by US. It specifies the latitude and longitude increments to sub-divide the rectangle into a grid-structure. The query begins by constraining the tweets to the bounding rectangle and those containing the hashtag ``Obama''. This step is executed efficiently by using the secondary R-tree index on the location attribute (from Figure~\ref{fig:feed_dataset_definition}).  The location of each qualifying tweet together with the origin of the bounding rectangle and the latitude and longitude increments (to specify the resolution of the grid) are given to the spatial-cell function. The function returns the grid cell that the tweet belongs to. Tweets are then grouped according to their containing grid cells and the count function is applied to each cell. The result can be used draw a heat map showing the relative volume of tweets over a selected geographic region.

\begin{figure}[htb]
%\vspace{-0.6cm}
\begin{aqlschema}
for $tweet in dataset ProcessedTweets
let $searchHashTag := "Obama"
let $leftBottom := create-point(33.13,-124.27)
let $rightTop := create-point(48.57,-66.18)
let $latResolution := 3.0
let $longResolution := 3.0
let $region := create-rectangle($leftBottom,$rightTop)
where spatial-intersect($tweet.location, $region) and
some $hashTag in $tweet.referred-topics
satisfies ($hashTag = $searchHashTag) 
group by $c := spatial-cell($tweet.location,
$leftBottom, $latResolution, $longResolution) with $tweet
return { "cell": $c, "count": count($tweet) }

\end{aqlschema}
\vspace{-0.4cm}
\caption{Spatial aggregation query over tweets generated in US and containing the hashtag “Obama”}
\label{fig:AQL_example_query}
\vspace{-0.2cm}
\end{figure}

\section{Data Feed Basics}
\label{sec:modeling_a_data_feed}

The AsterixDB  query language (AQL) has built-in support for data feeds. In this section, we describe how an end-user may model a data feed and have its data be persisted and indexed into an AsterixDB dataset. 
\vspace{-1mm}
\subsection{Collecting Data: Feed Adaptors} 
\label{sub-sec:collecting_data}
The functionality of establishing a connection with a data source, receiving, parsing and translating data into ADM records (for storage inside AsterixDB) is contained in a feed \emph{adaptor}. A feed adaptor is simply an implementation of an interface and its details are specific to a given data source. An adaptor may optionally be given configuration parameters as required to establish connection with the datasource and to configure runtime behavior. 
Depending upon the data transfer protocol/APIs offered by the data source, a feed adaptor may operate in a \emph{push} or a \emph{pull} mode. Push mode involves just one initial request (handshake) by the adaptor to the data source for setting up the connection and providing any protocol-specific parameters. Once a connection is established, the data 
source ``pushes'' data to the adaptor without any subsequent requests by the adaptor. In contrast, when operating in a pull mode, to receive data, the adaptor makes a separate request each time. 

AsterixDB currently provides built-in adaptors for popular data sources---Twitter, CNN, and RSS feeds. We are in process of expanding the set to cover other popular data sources. AsterixDB additionally provides a generic socket-based adaptor that can be used to ingest data that is directed at a prescribed socket.  Figure~\ref{fig:example_feeds} illustrates the use of built-in adaptors in AsterixDB to define a pair of feeds.  The \emph{TwitterFeed} contains tweets that contain the word ``Obama''. As configured, the adaptor will make a request for data every minute. The \emph{CNNFeed} will consists of news articles that are related to any of the topics that are specified as part of configuration.   

\vspace{-7mm}
\begin{figure}[htb]
\begin{aqlschema}

create feed TwitterFeed using TwitterAdaptor 
("api"="pull", "query"="Obama", "interval"=60);

create feed CNNFeed using CNNAdaptor 
("topics"="politics, sports");

\end{aqlschema}
\vspace{-0.5cm}
\caption{Defining a feed using some of the built-in adaptors in AsterixDB}
\label{fig:example_feeds}
%\vspace{-0.2cm}
\end{figure}

It is possible that the protocol for data exchange between the external source and the adaptor allows transfer of data in parallel across multiple channels. The degree of parallelism in receiving data from an external source is determined by the feed adaptor in accordance with the data exchange protocol. The TwitterAdaptor uses a single degree of parallelism whereas the CNNAdaptor  uses a  degree of parallelism  determined by the cardinality of the set of topics that is passed as configuration. Corresponding to each topic (politics, sports etc) is an RSS feed that is fetched by an individual instance of CNNAdaptor. Multiple instances of a feed adaptor may run as parallel threads on a single machine or on multiple machines. 

\subsection{Pre-Processing Collected Data}
A feed definition may optionally include the specification of a user-defined function that needs to be applied to each feed record prior to persistence. Examples of pre-processing might include adding/removing attributes, filtering out unwanted feed records, sampling, sentiment analysis or feature extraction for content-based classification. AsterixDB provides built-in support for creating user-defined functions (UDFs) in AQL or in programming languages like Java. 

The tweets collected by the TwitterAdaptor (Figure~\ref{fig:example_feeds}) conform to the \emph{RawTweet} datatype (Figure~\ref{fig:feed_data_specification}). The processing required in transforming a collected tweet to its lighter version (of type \emph{ProcessedTweet}) involves extracting hash tags\footnote{Hash tags are words that begin with a \#. In Twitter's jargon, these represent the topics associated with the tweet.} (if any) in a tweet and collect them under the referred-topics attribute for the tweet. This can be expressed as an AQL function. More sophisticated extract and collect pre-processing might require implementation in a programming language like Java. As an example, the CNNAdaptor (Figure~\ref{fig:example_feeds}) outputs records that each contain the fields---(item, link, description). The \emph{link} field provides the URL of the news article on the CNN website. Parsing the HTML source provides additional information such as  tags, images and outgoing links to other related articles. The extracted information could then be added to each record as additional fields to form an augmented version prior to persistence. 

The pre-processing function for a feed is specified using the \emph{apply function} clause at the time of creating the feed. This is illustrated in Figure~\ref{fig:example_feeds_udf}.

\begin{comment}Please note that the return type of the function associated with a feed must conform to the datatype of the target dataset where the feed is to be persisted. \end{comment}

\begin{figure}[htb]
\vspace{-0.6cm}
\begin{aqlschema}

create feed ProcessedTwitterFeed using TwitterAdaptor 
("api"="pull", "query"="Obama", "interval"=60)
apply function addHashTags;

create feed ProcessedCNNFeed using CNNAdaptor 
("topics"="politics, sports")
apply function extractInfoFromCNNWebsite;

\end{aqlschema}
\vspace{-0.4cm}
\caption{Defining a feed that involves pre-processing of collected data}
\label{fig:example_feeds_udf}
\end{figure}

A feed adaptor and a UDF act as \emph{pluggable} components that contribute towards providing a generic model for feed ingestion and help address challenge C1 from in Section~\ref{subsec:challenges}. By providing implementation of prescribed interfaces, the internal details of data feed management are abstracted from 
the end-user. These pluggable components can be packaged and installed as part of an AsterixDB library and subsequently be used in AQL statements.

\subsection{Building a Cascade Network of Feeds}
Multiple high-level applications might be driven by a data feed. Each such application might perceive feed data in a different way and require the arriving data to be processed and/or persisted differently. Building a separate flow of data from the external source for each application is wasteful of resources. Moreover the pre-processing or the transformation required by each application might overlap and could be done in an incremental fashion to avoid redundancy. A single flow of data from the external source then drives multiple applications. To achieve this, we introduce the notion of \emph{primary} and \emph{secondary} feeds in AsterixDB that help address challenge C2 from Section~\ref{subsec:challenges}.

A feed in AsterixDB is considered to be a \emph{primary} feed if it gets its data from an external data source. The records contained in a feed (subsequent to any pre-processing) are directed to an AsterixDB dataset. Additionally/alternatively, these records can be used to derive other feed(s) known as \emph{secondary} feed(s). A secondary feed is similar to its parent feed in every other aspect; it can have an associated UDF to allow for any subsequent processing, can be persisted into a dataset and/or be made to derive other secondary feeds to form a \emph{cascade network}. A primary feed and a dependent secondary feed mimic a parent-child relationship and form a hierarchy. To build an example, Figure~\ref{fig:example_feeds_secondary} shows the AQL statements that redefine the feeds---\emph{ProcessedTwitterFeed} and \emph{ProcessedCNNFeed}---in terms of their respective parent feeds from Figure~\ref{fig:example_feeds}.
\begin{figure}[htb]
\vspace{-0.6cm}
\begin{aqlschema}

create secondary feed ProcessedTwitterFeed from
feed TwitterFeed apply function addHashTags;

create secondary feed ProcessedCNNFeed from
feed CNNFeed apply function extractInfoFromCNNWebsite;

\end{aqlschema}
\vspace{-0.4cm}
\caption{Defining a secondary feed}
\label{fig:example_feeds_secondary}
\end{figure}

\subsection{Lifecycle of a Feed}
A feed is a \emph{logical} concept and is brought to life (i.e. its data flow is initiated) \emph{only} when it is \emph{connected} to a dataset using the \emph{connect feed} AQL statement (Figure~\ref{fig:feed_lifecycle}). Subsequent to a connect feed statement, the feed is said to be in the \emph{connected} state. Multiple feeds can simultaneously be connected to a dataset such that the dataset represents the union of the connected feeds. In a possible but unlikely scenario, a feed may also be simultaneously connected to different datasets. 
Note that connecting a secondary feed does not require the parent feed (or any ancestor feed) to be in the \emph{connected} state. The order in which feeds that are related in hierarchy are connected to the respective datasets is not important. Furthermore, additional (secondary) feeds can be added to an existing hierarchy and connected to a dataset at any time without interrupting the flow of data along a connected ancestor feed. 

\begin{figure}[!h]
  \includegraphics[width=90mm,height=45mm]{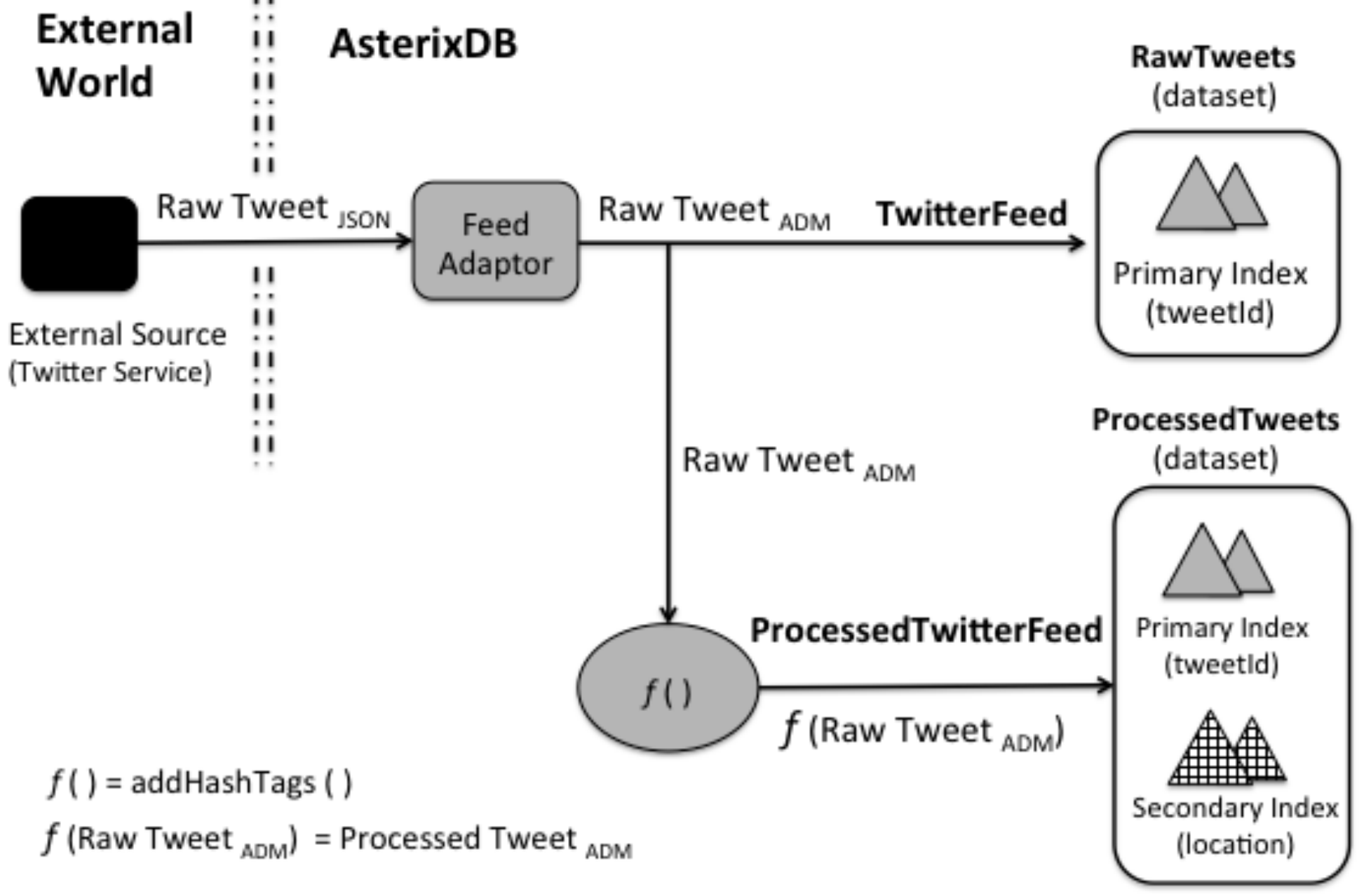}
  \caption{Logical view of the flow of data from external data source into AsterixDB datasets}
  \label{fig:primary_secondary_feed_flow}
\end{figure}

The \emph{connect feed} statement in Figure~\ref{fig:feed_lifecycle} directs AsterixDB to persist the \emph{ProcessedTwitterFeed} feed in the \emph{ProcessedTweets} dataset. If it is required (by the high-level application) to retain the raw tweets obtained from Twitter, end-user may additionally choose to connect  \emph{TwitterFeed} to a (different) dataset. Having a set of primary and secondary feeds offers the end-user the flexibility to do so. Let us assume that the high-level application needs to persist \emph{TwitterFeed} and that, to do so, the end-user makes use of  the \emph{connect feed} statement.  A logical view of the continuous flow of data as established on connecting the feeds to their respective target datasets is shown in Figure~\ref{fig:primary_secondary_feed_flow}. 
\vspace{-0.4cm}
\begin{figure}[!ht]
\begin{aqlschema}
connect feed ProcessedTwitterFeed to 
dataset ProcessedTweets;

disconnect feed ProcessedTwitterFeed from 
dataset ProcessedTweets;
\end{aqlschema}
\vspace{-0.4cm}
\caption{Managing the lifecycle of a feed}
\label{fig:feed_lifecycle}
\end{figure}

Contrary to the \emph{connect feed} statement, the flow of data from a feed into a dataset can be terminated explicitly by use of  the \emph{disconnect feed} statement (Figure~\ref{fig:feed_lifecycle}). Note that disconnecting a feed from a particular dataset neither interrupts the flow of data from the feed to any other dataset(s) nor does it impact other connected feeds in the lineage.

%\vspace{-0.2cm}
\begin{comment}
The notion of a primary and a secondary feed allows the end-user to derive a feed from an existing feed 
and form a cascade network with data flowing from one feed to another and so on and so forth. This ability coupled with the flexibility to connect each feed in a cascade network to a different dataset helps in 
addressing the \emph{challenge}---C2---that was described in Section~\ref{subsec:challenges}. Figure~\ref{fig:primary_secondary_feed_flow} illustrates the \emph{Fetch-Once Compute-Many} model where a record (tweet) is retrieved from the external source (Twitter) once but is pushed along multiple paths to be processed and persisted differently. 
\end{comment}

\subsection{Policies for Feed Ingestion}
\label{sec:policies_feed_ingestion}
Multiple feeds may be concurrently operational on a given AsterixDB cluster, each competing for resources (CPU cycles, network bandwidth, disk IO) to maintain pace with their respective data sources. A data management system must be able to manage a set of concurrent feeds and take dynamic decisions related to allocation of resources, resolving resource bottlenecks and handling of application/hardware failures. Moreover, a data management system must do so in a generic way without being tightly coupled with a restricted set of data sources or applications. In this sub-section, we discuss the approach adopted in AsterixDB.    

Each feed has its own set of constraints influenced largely by the nature of its data source and the application(s) that intend to process the ingested data. Consider an application that intends to discover the trending topics on Twitter by analyzing the feed---\emph{ProcessedTwitterFeed}. Losing a few tweets may be acceptable. In contrast, when ingesting from a data source that provides a click-stream of ad clicks, losing data translates to a loss of revenue for an application that generates revenue by charging advertisers per click. In a resource-constrained environment, records from a feed could be spilled to disk and processed later. Additionally/alternatively a threshold number of records (expressed as a fraction of total ingested records) could be discarded altogether to limit the demand of resources. Application(s) may optionally require monitoring and reporting of metrics associated with the feed and dictate the kind of failures that the feed should survive during its lifetime.  Such aspects associated with feed ingestion are expressed as a collection of parameters and associated values that together form an ingestion policy.

A custom policy can be created by choosing an appropriate value for each policy parameter. Alternatively, parameter values from an existing policy can be tweaked to form a new policy. A few important policy parameters are described in Table~\ref{BuiltinPolicies}. AsterixDB comes with a set of built-in policies (\emph{Basic, Monitored, Fault-Tolerant}). The \emph{Basic} policy does not provide support for handling of failures. The \emph{Monitored} policy extends the \emph{Basic} policy by enforcing collection 
of metrics (rate of flow of data, CPU utilizations etc.) to be collected for logging/reporting. The \emph{Fault-Tolerant} policy is appropriate when the feed needs to survive failures. For an elaborate list of policy parameters, the reader is referred to \cite{AsterixDB}. The desired feed ingestion policy is specified as part of the connect feed statement as shown in Figure~\ref{fig:feed-policy-specification}. The Monitored policy is chosen as the default if a policy is not specified explicitly. Note that Figure~\ref{fig:feed-policy-specification} shows an example where a primary feed (\emph{TwitterFeed}) and a dependent secondary feed (\emph{ProcessedTwitterFeed}) are both connected using a common policy (Basic), but this is not a  requirement. The ability to form a custom policy allows the runtime behavior to be customized 
as per the specific needs of the high-level application(s) and helps address challenge C1 from Section~\ref{subsec:challenges}. 

\begin{figure}[htb]
\vspace{-0.2cm}
\begin{aqlschema}
connect feed TwitterFeed to dataset RawTweets
using policy Basic;

connect feed ProcessedTwitterFeed to 
dataset ProcessedTweets using policy Basic;
\end{aqlschema}
\vspace{-0.5cm}
\caption{Specifying the ingestion policy for a feed}
\label{fig:feed-policy-specification}
\end{figure}
\vspace{-0.5mm}

\begin{center}
   \begin{table} 
       \small        
        \caption{A Few Important Policy Parameters}
         \begin{tabular} [!ht] { | p{2.5cm} | p{5.5cm} |} \hline
	        Policy &  Description \\ \hline \hline
         	excess.records.spill & Set to true if records that cannot be processed by an operator for want of resources should be persisted to (local) disk and processed later when resources become available.  A false value causes such records to be discarded altogether.
         	 \\ \hline
			recover.soft.failure &  Set to true if the feed must attempt to survive any runtime exception, else an exception causes an early termination the feed.\\ \hline
			recover.hard.failure & Set to true if the feed must attempt to survive a hardware
			 failures (loss of single/multiple AsterixDB node(s)), else a hardware failure causes the feed to terminate. \\ \hline
  \end{tabular}
  \label{BuiltinPolicies}
  \end{table} 
\end{center}
\vspace{-1.0cm}

\section{Runtime for Feed Ingestion}
\label{sec:runtime_for_feed_ingestion}
So far we have described, at a logical level, the user model and built-in support in AQL that enables the end-user to define a feed and manage its lifecycle. In this section, we delve into the physical aspects and implementation details involved in building and managing the flow of data when a feed is connected to a dataset. 

\subsection{Components of Runtime}
In processing a connect feed statement, the AQL compiler retrieves the definitions of the involved components---feed, adaptor, function, policy and the target dataset from the AsterixDB Metadata. The  compiler translates a \emph{connect feed} statement into a Hyracks job that is subsequently scheduled to run on an AsterixDB cluster. The resulting dataflow is referred to as a feed ingestion pipeline. A \emph{data operator} forms a major building block of an ingestion pipeline and
is useful in executing custom logic on partitions of input data to produce partitions of output data. It may employ parallelism in consuming input by having multiple instances that run in parallel across a set of nodes in an AsterixDB cluster. \emph{Data connectors} repartition operators' outputs to make the newly produced partitions available at the consuming operator instances. In addition, an ingestion pipeline provides \emph{feed joints} at specific locations. A \emph{feed joint} is like a network tap and provides access to the data flowing along a pipeline. It offers a subscription mechanism and allows the data to be routed simultaneously along multiple paths to individual subscribers for building a cascade network. 

A feed ingestion pipeline involves 3 stages---\emph{intake}, \emph{compute} and \emph{store}. The intake stage involves creating an instance of the associated feed adaptor, using it to initiate transfer of data and transforming into ADM records. If the feed has an associated pre-processing function, it is applied to each feed record as part of the \emph{compute} stage. Subsequently, as part of the \emph{store} stage, the output records from the preceding \emph{intake/compute} stage are put into the target dataset and secondary indexes\footnote{Secondary indexes in AsterixDB are partitioned and co-located with the corresponding primary index partition. Insert of a record into the primary and any secondary indexes uses write-ahead logging and offers ACID semantics.} (if any) are updated accordingly. Each stage is handled by a specific data-operator, hereafter referred to as an \emph{intake}, \emph{compute}, and \emph{store} operator respectively. 

\begin{figure}[!ht]
  \includegraphics[width=75mm,height=30mm]{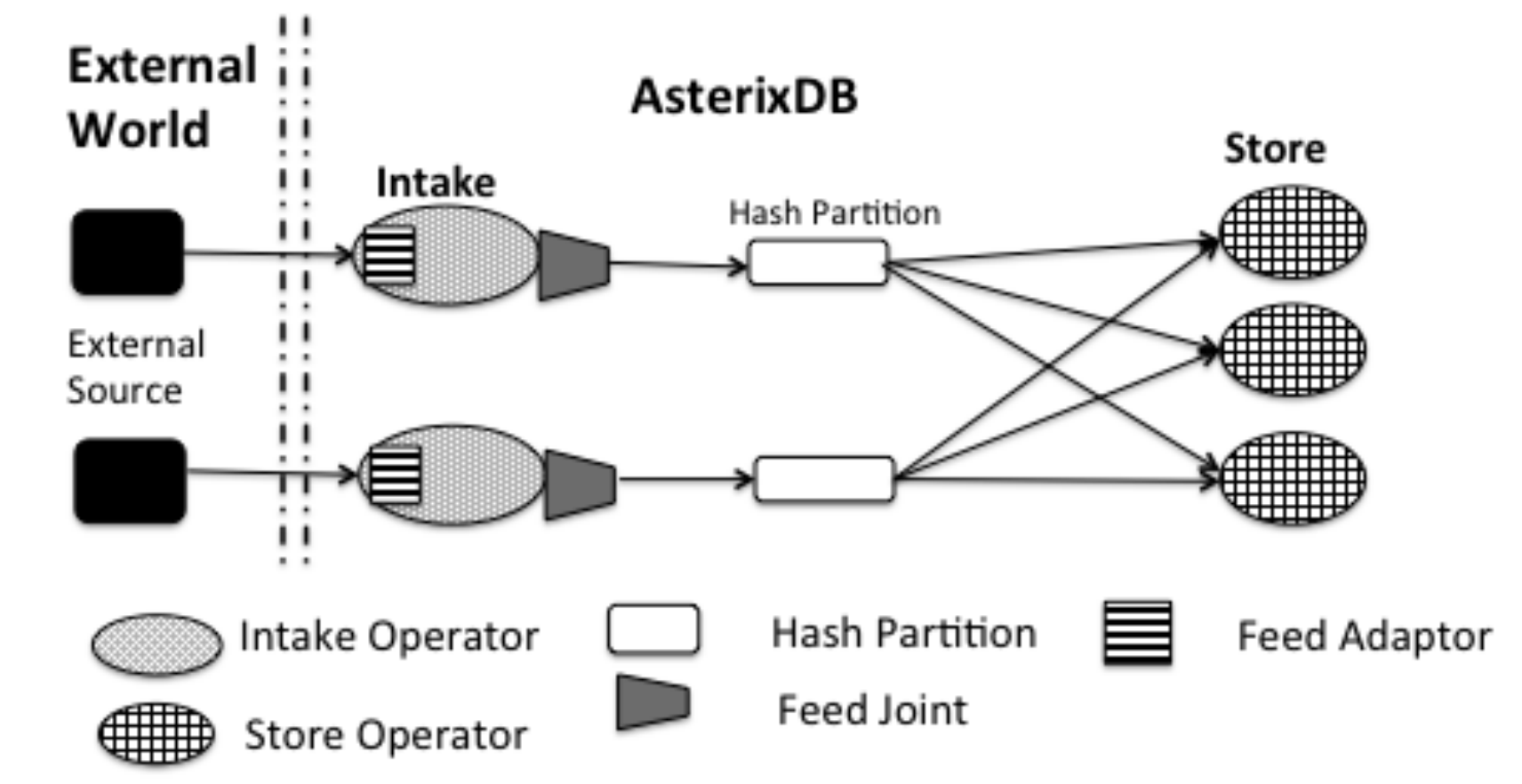}
  \caption{An example feed ingestion pipeline}
  \label{fig:secondary_feed_ingestion}
\end{figure}

Next, we describe how operators, connectors and joints are assembled together to  construct a feed ingestion pipeline. 
\vspace{-0.4cm}
\begin{figure}[htb]
%\vspace{-0.2cm}
\begin{aqlschema}
connect feed CNNFeed to dataset RawArticles;

connect feed ProcessedCNNFeed to 
dataset ProcessedArticles;
\end{aqlschema}
\vspace{-0.3cm}
\caption{Example AQL statements}
\label{fig:constructing_ingestion_pipeline}
\end{figure}
The first statement in Figure~\ref{fig:constructing_ingestion_pipeline} connects the primary feed---CNNFeed. As determined by the cardinality of the set of topics specified as configuration, the feed involves the use of a pair of instances of the CNNFeedAdaptor. Each adaptor instance is created and managed by an instance of the \emph{intake} data operator. As \emph{CNNFeed} does not involve any pre-processing, the output records from each adaptor instance thus constitute the feed. These are then partitioned across a set of \emph{store} operator instances by the hash-partitioning data-connector. The constructed pipeline is shown in Figure~\ref{fig:secondary_feed_ingestion}. Notice that a feed joint is located at the output of each intake operator instance. In general, a feed joint is placed at the output side of an operator instance that produces records that form the feed. In the case when a feed involves pre-processing, a feed joint is placed at the output of each of the \emph{compute} operator instances.   

The second statement in Figure~\ref{fig:constructing_ingestion_pipeline} connects the secondary feed---\emph{ProcessedCNNFeed}. By definition, the feed can be obtained by subjecting each record from the \emph{CNNFeed} to the associated UDF (\emph{extractInfoFromCNNWebsite}). 
In general, given that $feed_{m+1}$ denotes the immediate child of $feed_{m}$, a child feed $feed_{i}$ can be obtained from an ancestor feed $feed_{k}$ $(k<i)$ by subjecting each record from $feed_{k}$ to the sequence of UDFs associated with each child feed $feed_{j}$ $(j=k+1,...,i)$. To minimize the additional processing involved in forming a child feed, it is desired to source the feed from its closest ancestor feed that is in the connected state. The feed joint(s) available along the feed ingestion pipeline of an ancestor feed are then used to access the flowing data and subject it to additional processing to form the desired feed. AsterixDB keeps track of the available feed joints and uses them in preference over a feed adaptor in sourcing a feed. 

\begin{figure}[!h]
 
  \subfigure[Flow of data along a cascade network when ProcessedCNNFeed is connected to a dataset] { 	  \includegraphics[width=75mm,height=40mm]{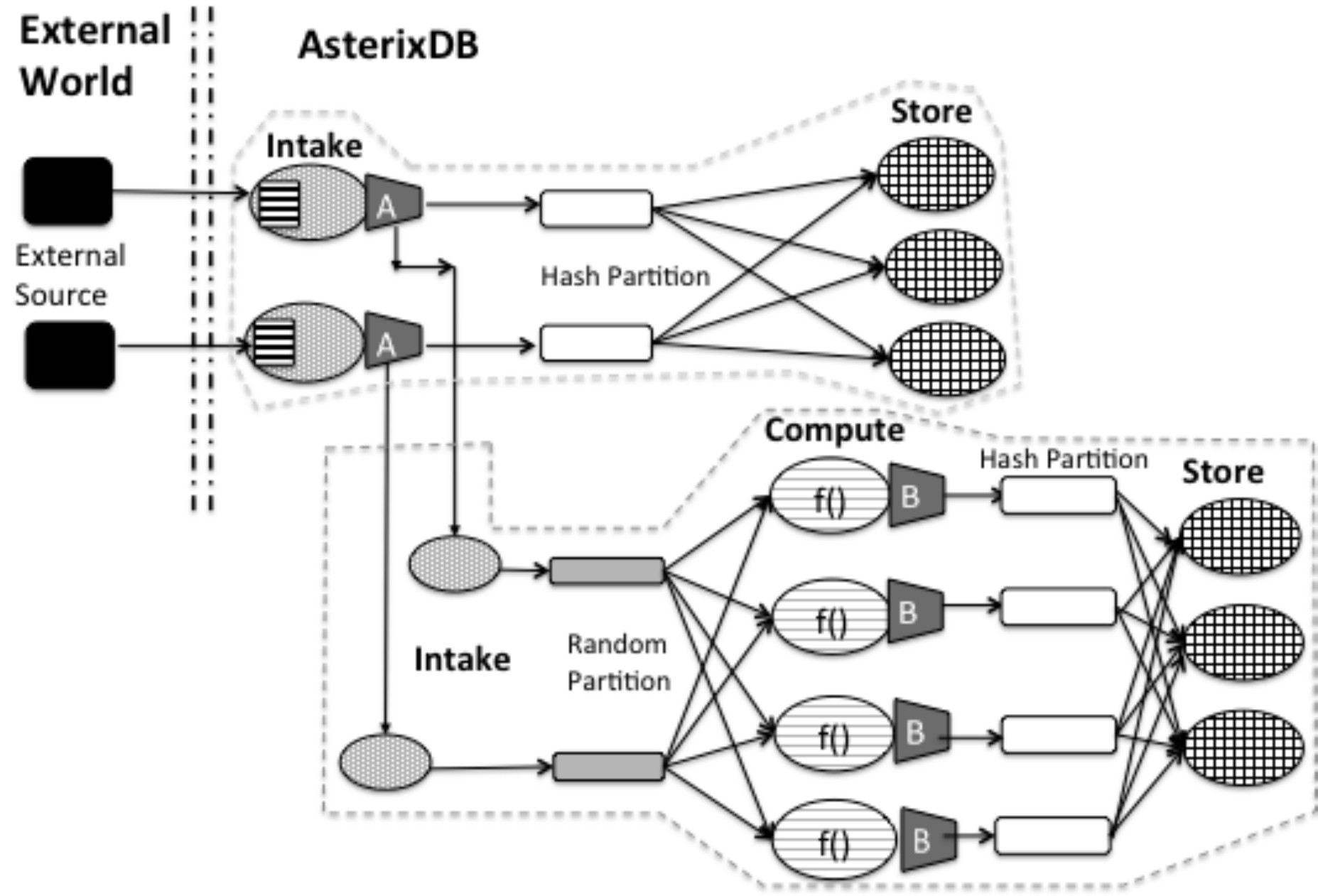}
  \label{fig:secondary_primary_feed_ingestion}
  }
  
  \subfigure[Flow of data subsequent to disconnection of CNNFeed from its target dataset] { 	  
  \includegraphics[width=75mm,height=40mm]{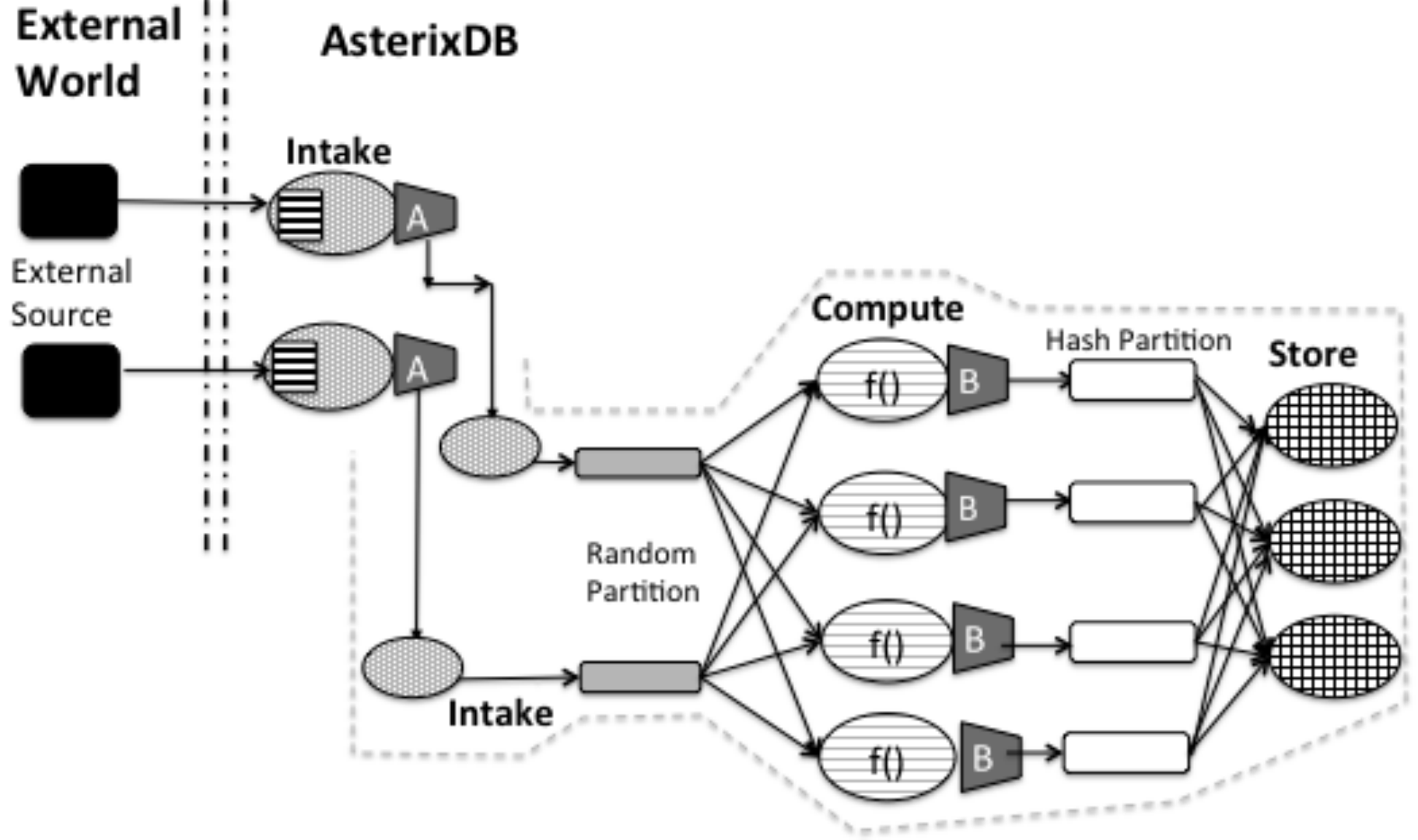}
  \label{fig:primary_only_feed_ingestion}
  }
  \caption[]{An example of a feed cascade network}
\vspace{-4mm}
\end{figure}

The ingestion pipeline for \emph{ProcessedCNNFeed} is shown in Figure~\ref{fig:secondary_primary_feed_ingestion}. The intake operation involves a pair of intake operators, each receiving CNNFeed records from the respective feed joints (kind A). Output records from the intake operator instances are randomly partitioned across a set of compute nodes (4 in the example pipeline) that apply the associated UDF to each record to produce the secondary feed---ProcessedCNNFeed. Subsequently the feed records are hash-partitioned across a set of store operator instances for persistence. The cascade network provides an additional set of feed joints (kind B) that provide access to CNNProcessedFeed. If at this stage, the end-user creates (and connects) a secondary feed that derives from \emph{ProcessedCNNFeed}, then its intake stage would involve receiving records from each of the 4 feed joints (kind B) located along the ingestion pipeline for \emph{ProcessedCNNFeed}. 

\begin{comment}
Note that data may not move at the same rate along the paths emanating from a feed joint. For e.g. application of an expensive function along a path can cause significant delays. The implementation of a feed joint ensures that a sluggish data path emanating from a given feed joint does not impact movement of data across other paths originating from the feed joint. At a feed joint, an arriving frame of data is deposited in a shared queue. The queue is processed independently by each subscriber. As and when a frame has been consumed by each subscriber, its A frame of data (contaifeed records) arriving at a feed joint is held in memory until is consumed by each subscriber.  
\end{comment}

It is worth noting that disconnecting a feed from a dataset does not necessarily remove the set of feed joints located along the ingestion pipeline. Referring to the data flow shown in Figure~\ref{fig:secondary_primary_feed_ingestion}, disconnecting CNNFeed at this stage removes the tail of the pipeline that includes the compute and store operator instances but retains the intake operator instances. This is because the feed joints (kind A) at the output of the intake operator instances, each have an existing subscriber that requires the output records to keep flowing in an uninterrupted manner. The resulting data flow is shown in Figure~\ref{fig:primary_only_feed_ingestion}.

\subsection{Scheduling a Feed Ingestion Pipeline} 
\label{subsec:SchedulingPipeline}
Scheduling a feed ingestion pipeline on a cluster requires determining the desired cardinality (degree of parallelism) of each operator and mapping each instance of an operator to an AsterixDB node. The location and cardinality constraints for the intake operator are determined by the feed adaptor. If no constraints are specified, AsterixDB chooses to run a single instance on a randomly chosen node. In contrast to the intake operator, the location (and cardinality) constraints for the store operator are pre-determined and derived from the nodegroup associated with the target dataset. Recall that the nodegroup of a dataset refers to the set of nodes that hold the partitions of the dataset. 

The compute operator is different from the intake/store operator as by definition, it can be placed at any AsterixDB node and offers no location or cardinality constraints. The partitioned parallelism employed at the compute and store stages helps the system ingest increasingly large volume of data. Additional resources (physical machines) can be added at the compute and/or store stage to scale out the system. This helps address challenge C5 from Section~\ref{subsec:challenges}. The appropriate degree of parallelism is dependent on the rate of arrival of data and the complexity associated with the UDF being applied at the compute stage. A lesser degree of parallelism can cause sluggish data movement while excessive parallelism can be wasteful of resources. To begin with, the cardinality at the compute stage is matched with that of the store stage to offer a similar degree of parallelism. However, as we describe in the following section, a feed ingestion pipeline needs to be monitored for resource bottlenecks and if required, needs to be re-structured in accordance with the demand for resources. 

\subsection{Managing a Feed Ingestion Pipeline} 
\label{subsec:MonitoringPipeline}
Data travels along a feed ingestion pipeline as fixed-size chunks known as \emph{frames} in Hyracks. Each frame contains a variable number of records. Resource constraints at the node hosting an operator may cause delays in processing of records. An expensive UDF and/or an increased rate of arrival of data may lead to an excessive demand for resources. Delays in processing by a downstream operator slows down the upstream operator as it is no longer able to send data downstream. This can potentially cascade upstream to parent operators as \emph{back-pressure} and `lock' the flow of data.  In this sub-section, we present the methodology adopted in AsterixDB for monitoring an ingestion pipeline for resource bottlenecks and taking corrective action where necessary (challenge C3 from Section~\ref{subsec:challenges}). 

An AsterixDB node runs as a Java process(JVM) that is configured with a limit to the amount of available memory. Besides supporting flow of data along feeds, an AsterixDB node also participates 
in execution of queries that involve aggregation, joining and sorting of data and manages its memory consumption. During normal operation, an operator instance in a feed ingestion pipeline uses a limited amount of memory as it stores input/output frames in reusable buffers. However, feed records may be pushed to AsterixDB at  a rate that is higher than what the ingestion pipeline may consume. The possibility of regulating the rate at the external source cannot be assumed. To prevent data loss, additional memory is required by operator instances to buffer the arriving records until the backlog of preceding records is cleared.  

Note that operators involved in a feed ingestion pipeline (\emph{intake}, \emph{compute} and \emph{store}) are reusable components that get employed in executing other AQL statements/queries. It is desirable to keep these operators unchanged so that they remain simple and generic for use elsewhere as part of other jobs. To transparently add monitoring and buffering capabilities, we wrap each participant operator with a \emph{MetaFeed} operator. The MetaFeed operator provides an interface identical to that offered by the underlying wrapped operator (hereafter referred to as the \emph{core} operator) but adds functionality to buffer the input records, monitor the rate of consumption by the core operator and take necessary action if the core operator is unable to process records at their arrival rate. Each worker node in an AsterixDB cluster hosts a \emph{Feed Manager}. Amongst the Feed Managers corresponding to each AsterixDB node, a \emph{leader} is chosen and referred to as the \emph{Super Feed Manager}. Each instance of the MetaFeed operator registers itself with the local Feed Manager. 

The input records to a MetaFeed operator are made to wait in fixed-size re-usable buffers until the core operator is able to process them. To govern memory allocation, each AsterixDB node hosts a Feed Memory Mananger (FMM) that is initialized with a global budget in terms of number of (fixed-size) buffers that may be allocated at any point in time. The MetaFeed operator monitors the rate of consumption of records by the core operator and may request for allocation of additional buffers to accommodate the arriving records. In a favorable case, the FMM responds with allocation of a limited (but configurable) number of buffers. An allocated buffer is returned to the FMM and the occupied  memory reclaimed when it is no longer required by the operator instance. In an unfavorable case, the request (for additional memory) can be turned down if it violates the global budget. The MetaFeed operator then reports this to the local Feed Manager as a \emph{stalled} state.  

On being notified of a \emph{stalled} state, the Feed Manager attempts a local resolution by spilling the continuously arriving records to disk (deferred processing) or discarding the records altogether. The precise behavior and the associated limits are controlled by the ingestion policy associated with the feed. The  resolution (spilling/discarding) initiated by the Feed Manager helps in localizing the congestion and prevents it from escalating upstream as \emph{back-pressure} that can eventually `lock' the flow of data along the pipeline. Recall that in the case of a cascade network,  multiple ingestion pipelines receive data from a common feed joint. A `locked' state of a pipeline puts excessive demand for resources at the feed joint and can impede the flow of data along other ingestion pipelines originating from the feed joint. Localizing the congestion helps to prevent such an undesirable situation. Note that if the ingestion policy does not permit spilling or discarding of records or the associated limits have been utilized, the Feed Manager is unable to take any further action. In such a scenario, the FeedManager notifies Super Feed Manager of the \emph{stalled} state. 

A Super Feed Manager is able to form a global view of the feed ingestion pipeline and locate all operators that are reporting such states. In addition, the Super Feed Manager also receives periodic report messages from all Feed Managers corresponding to each node in an ingestion pipeline. Each report message includes statistics related to the CPU and disk utilizations and the rate of inflow/outflow of data at their AsterixDB node.  The Super Feed Manager can then restructure (alter the degree of parallelism for a stage) and/or relocate the ingestion pipeline of the feed, and may even resort to re-sizing the cluster by adding additional nodes from a pre-configured pool of spare machines. The methodology for forming the corrective action and subsequent evaluation and modifications (if required) is not discussed further as it is part of the ongoing work.

%\subsection{Resource Management}
%\label{subsec:resource_management}
\begin{comment}
An AsterixDB node runs as a Java process(JVM) that is configured with a limit to the amount of available memory. Besides supporting flow of data along feeds, an AsterixDB node also participates 
in execution of queries that involve aggregation, joining and sorting of data and manages its memory consumption. During normal operation, an operator instance in a feed ingestion pipeline uses a limited amount of memory as it stores input/output frames in reusable buffers. However, feed records may be pushed to AsterixDB at  a rate that is higher than what the ingestion pipeline may consume. The possibility of regulating the rate at the external source cannot be assumed. To prevent data loss, additional memory is required by the intake operator to hold the arriving records in memory-resident buffers until the preceding records are processed and sent downstream.  
\end{comment}
\vspace{-0.2cm}

\section{Fault Tolerant Feed Ingestion}
\label{sec:fault_tolerant_feed_ingestion}

Feed ingestion is a long running task and is bound to encounter hardware failure(s) as it continues to run on a cluster of machines. Furthermore, parts of a feed ingestion pipeline include pluggable user-provided modules (feed adaptor and a pre-processing function) that may cause soft failures in the form of runtime exceptions.  Sources of an exception include unexpected data format, unexpected null values for an attribute, or simply inherent bugs in the user-provided source code that show up for certain kind(s) of data values. We categorize failures occurring from processing of data as software failures, and those arising from loss of a physical machine (due to a disk,network or power failure) as hardware failures. In this section, we describe how a feed may recover from software and hardware failures and particularly address the challenge C4 from Section~\ref{subsec:challenges}. The kind of failures a feed is required to survive is determined by the associated ingestion policy. 

\subsection{Handling Software Failures}
A runtime exception encountered by an operator in processing an input record in a typical insert setting carries non-resumable semantics and causes the dataflow to cease. It is essential to guard the feed pipeline from such exceptions by executing each operator in a sandbox-like environment. The MetaFeed operator (introduced in Section~\ref{subsec:MonitoringPipeline}) acts as a shell around each operator to provide such an environment. Recall that the operator that is wrapped is referred to as the \emph{core} operator.
The runtime of a core operator receives input data as a sequence of frames each comprising of records. An exception thrown by the core operator in processing an input record is caught by the wrapping MetaFeed operator. The MetaFeed operator slices the original input frame to form a subset frame that excludes the processed records and the exception generating record.  The subset frame is then passed to the core-operator which continues to process input frames and has, in effect skipped past the exception-generating record. 

The MetaFeed operator provides different options for handling/logging the exception. At minimum, the exception and the causing record are appended to the standard AsterixDB error log file. Alternatively, the information may also be persisted into a dedicated AsterixDB dataset. The logging support for a feed is configured as part of the ingestion policy. In a possible scenario, every record may result in a similar exception; a situation indicative of a bug. A repeated cycle of handling/logging of exception in such a case is wasteful of resources. To avoid such a situation, a feed ingestion policy can be configured with an upper bound on the number of consecutive records that can be ``skipped'' by an operator. Upon reaching the limit, an exception raised on the next record causes the faulty feed to end.
\\
\subsection{Handling Hardware Failures}
In this section, we describe the mechanism by which AsterixDB handles the loss of one or more of the AsterixDB nodes involved in a feed ingestion pipeline. Corresponding to the operation being performed, a node is referred to as an \emph{Intake}, \emph{Compute} or a \emph{Store} node. A node may simultaneously act as an intake, compute or a store node for one or more feeds. Recovering the loss of a node requires the failed node to be substituted by a different node and the feed ingestion pipeline to be rescheduled to involve the substitute node. Any other node in an AsterixDB cluster may be chosen as a substitute. Alternatively a new node (from a pre-configured pool of spares) can be dynamically added to the cluster to act as a substitute. 

\begin{figure}[!h]
  \includegraphics[width=90mm,height=50mm]{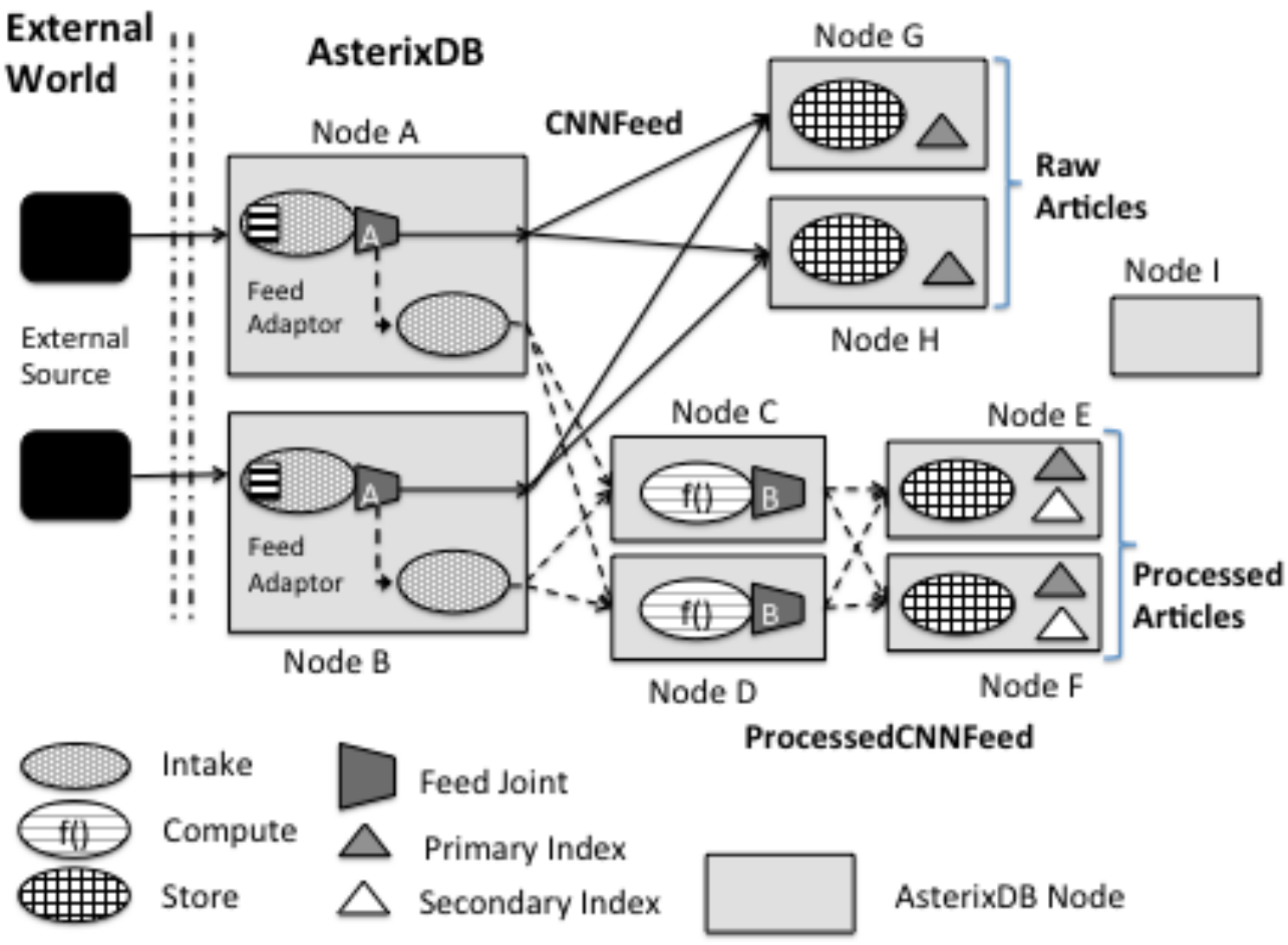}
  \caption{An example feed ingestion dataflow for describing the fault tolerance protocol}
  \label{fig:feed_flow_fault_tolerance}
\end{figure}

To illustrate different failure scenarios, we revisit our example data flow shown in Figure~\ref{fig:secondary_primary_feed_ingestion} and form a simplified version that uses a lower degree of parallelism at the compute and storage tier and yet allows us to describe all aspects of the fault-tolerance protocol. The simplified example data flow executes on a 10 member AsterixDB cluster (nodes A--I as shown in Figure~\ref{fig:feed_flow_fault_tolerance} and an additional \emph{master} node). In this particular data flow, node I is not used initially.  
To be considered alive, each node is required to send periodic heartbeats to the master node. 

In a feed ingestion pipeline, the data flowing between the operators is pipelined with all operators executing in parallel.  When an AsterixDB node fails, its operator instances are lost and are referred to as \emph{dead} instances. Subsequently, the other operator instances in the same pipeline that are running on other (alive) AsterixDB nodes are notified of the pipeline failure. On being notified, the operator instance   
saves the set of pending frames from its input/output queues by giving it to the local Feed Manager. In addition, the operator has an option to save state information that may help in resuming operation once the pipeline is rescheduled. The instance terminates itself and is then referred to as a \emph{zombie} instance, as it is not alive as a JVM thread but has runtime state information available for future retrieval.

There are two scenarios where an operator instance does not transit to the \emph{zombie} state.
First, if the output from an operator instance is being routed along multiple paths via a local feed joint, the operator instance must continue to output data to maintain a continuous flow along the dependent ingestion pipeline(s). Second, an intake operator must continue to live to maintain the flow of data from the data source, as an interrupted data flow could lead to irrecoverable data loss\footnote{A data source may allow retrieval of past records from a live feed. In this case, feed records not fetched during recovery phase can be fetched later.}.

Subsequently, the feed ingestion pipeline enters a \emph{recovery} phase.  A recovering feed ingestion pipeline is re-constructed with identical operators and feed joints but the location for each operator instance in the newly constructed pipeline is chosen carefully. An operator instance is co-located with its zombie instance from the previous failed execution, if possible (node is still available). This allows the new operator instance to collect any saved state (left by its \emph{zombie} instance) from the local Feed Manager. An operator instance that has a \emph{dead} instance from the previous execution can be scheduled to run at any AsterixDB node. Note that any state information from a dead instance is lost. An intake operator instance is co-located with the corresponding \emph{live} instance from the previous execution.  The functionality of registering with the Feed Manager and saving/retrieving any state  across failures is provided by the \emph{MetaFeed} operator that wraps around a core operator. Next, we consider different example failure scenarios and describe the recovery phase.

\textbullet\ External Data Source Failure: The external data source runs on just another physical machine outside the AsterixDB cluster. A power/disk failure at the external machine or lost network connectivity would interrupt the regular flow of data. For example, Twitter or CNN as a data source may experience an outage.  AsterixDB remains agnostic of such failures as these are transparently handled by the adaptor. An adaptor may resort to reconnecting after a wait or connecting to a different server/machine offered as part of the agreed protocol. However, if the adaptor is not in a position to continue, it must convey this to AsterixDB, in which case AsterixDB terminates the feed and relinquishes any involved resources. 
\begin{comment}On the contrary, if the adaptor is able to recover from the failure (via reconnecting, switching to a different source node etc) the feed continues to live. AsterixDB then remains agnostic of any observed issues faced by the adaptor, though it may see a transient drop in the feed throughput.
\end{comment}
   
\begin{figure}[!ht]
  \includegraphics[width=85mm,height=45mm]{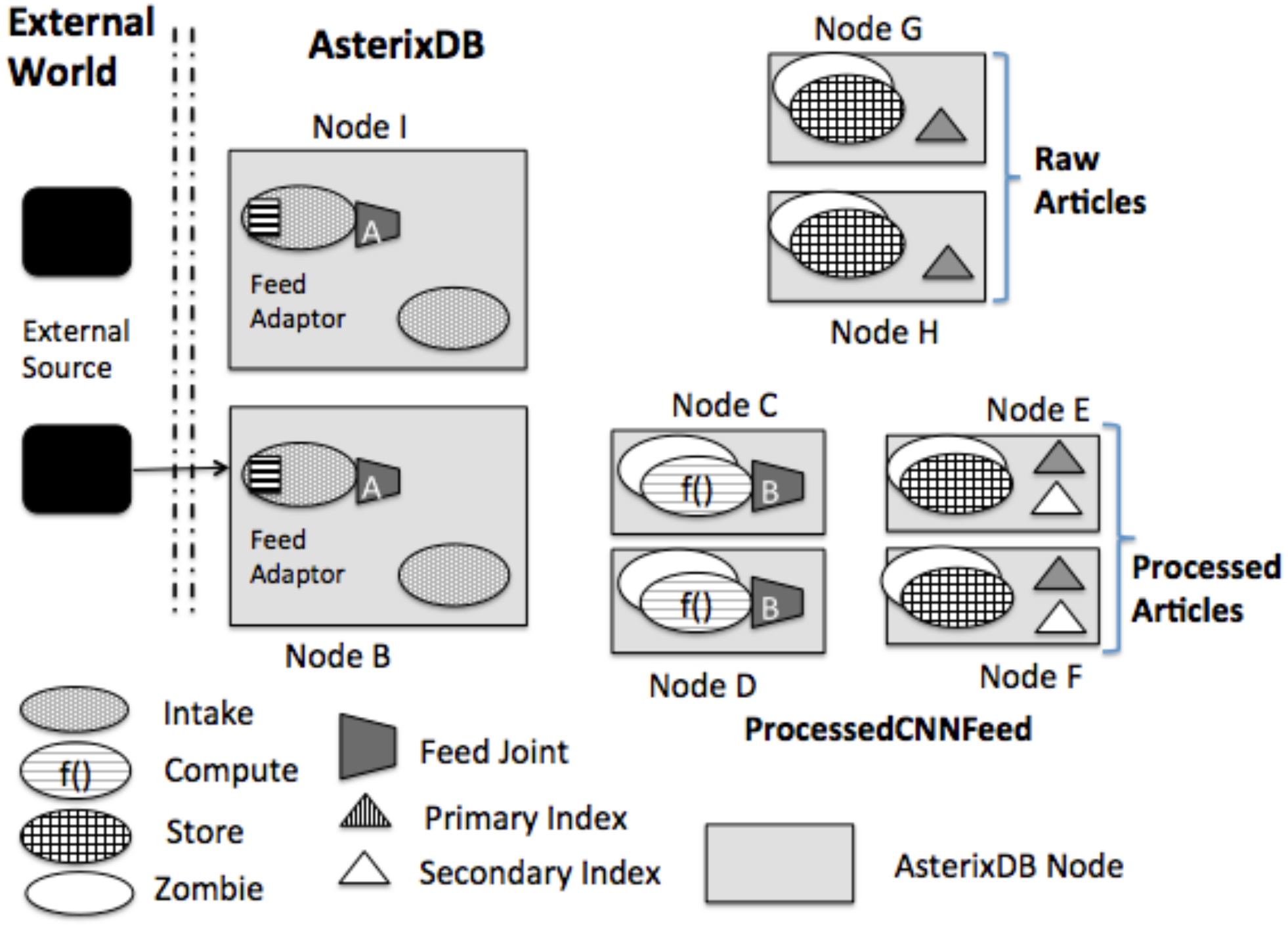}
  \vspace{-4mm}
  \caption{Intake Node Failure: Transient state during recovery phase.}
  \label{fig:IntakeFailureIntermediate}
\end{figure}

\textbullet\ Intake Node Failure: Referring to our example data flow of Figure~\ref{fig:feed_flow_fault_tolerance}, we assume the loss of node A while the feeds---CNNFeed and ProcessedCNNFeed---are active. Both feeds loose intake operator instance at Node A. The notified compute and store operator instances executing at nodes C---H save their state with the local Feed Manager and enter the \emph{zombie} state. Node I, which is idle, is the preferred choice to substitute node A. Figure~\ref{fig:IntakeFailureIntermediate}  shows a transient state during the recovery phase. As shown, a new intake operator instance is placed at node I and new (compute, store) operator instances at nodes C--H replace the zombie instances. Node I has substituted node A but is yet to re-establish connection with data source. The intake operator instance at node B continues to receive data from the external data source as before, but buffers it until the pipeline is restored. 
Subsequently, the following happens. Each of the (compute, store) operator instances at nodes C--H retrieve the saved state from their local feed managers. The new intake operator instance at node I establishes a new connection with the external source. The intake operator instance at node B forwards the buffered records and resumes normal operation. 

\begin{figure}[!h]
  \subfigure[Transient state during recovery phase.] { 	  
  \includegraphics[width=85mm,height=35mm]{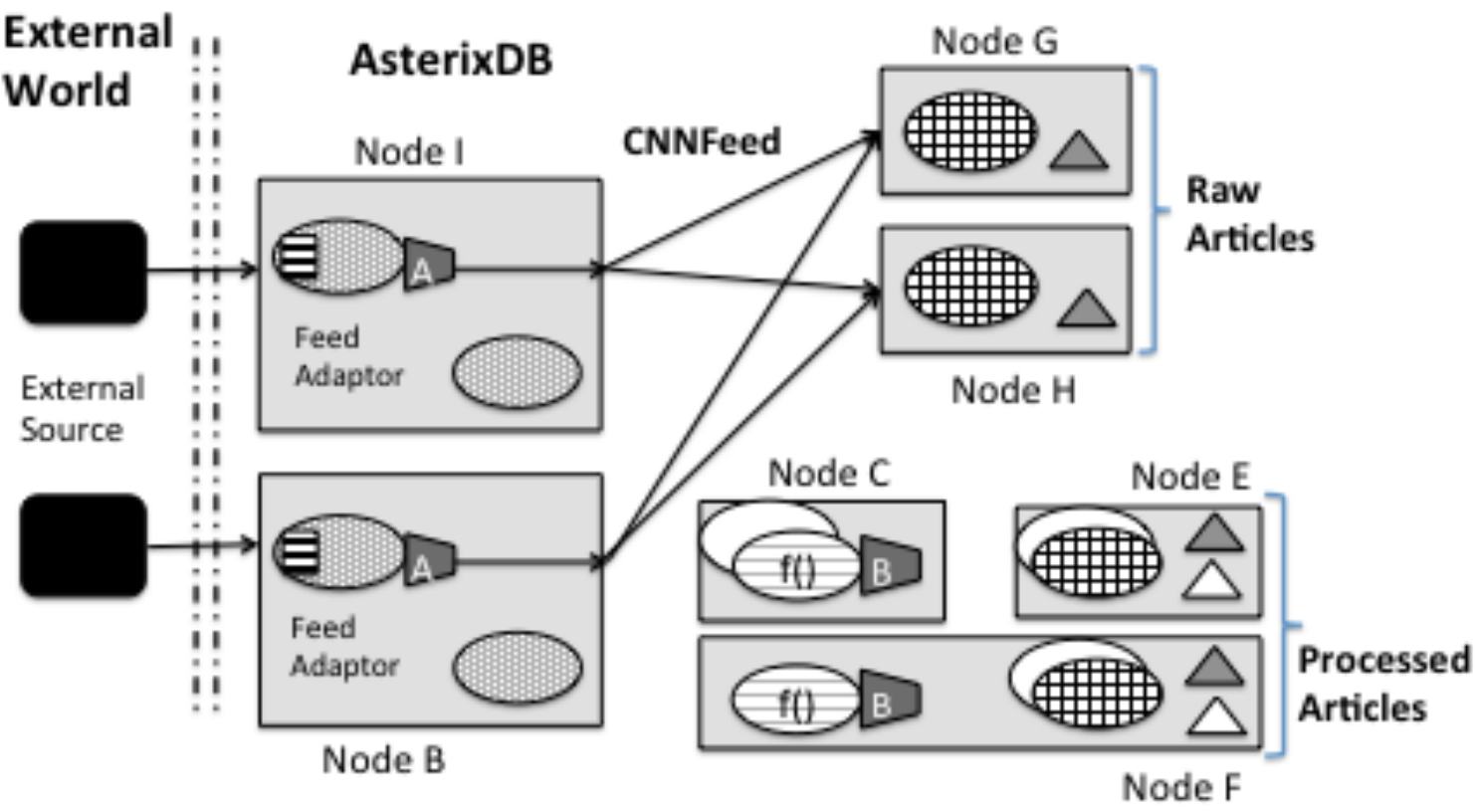}
  \label{fig:compute_node_failure_intermediate_stage}
  }
  
  \subfigure[Restructured pipeline post recovery.] { 	  
  \includegraphics[width=85mm,height=45mm]{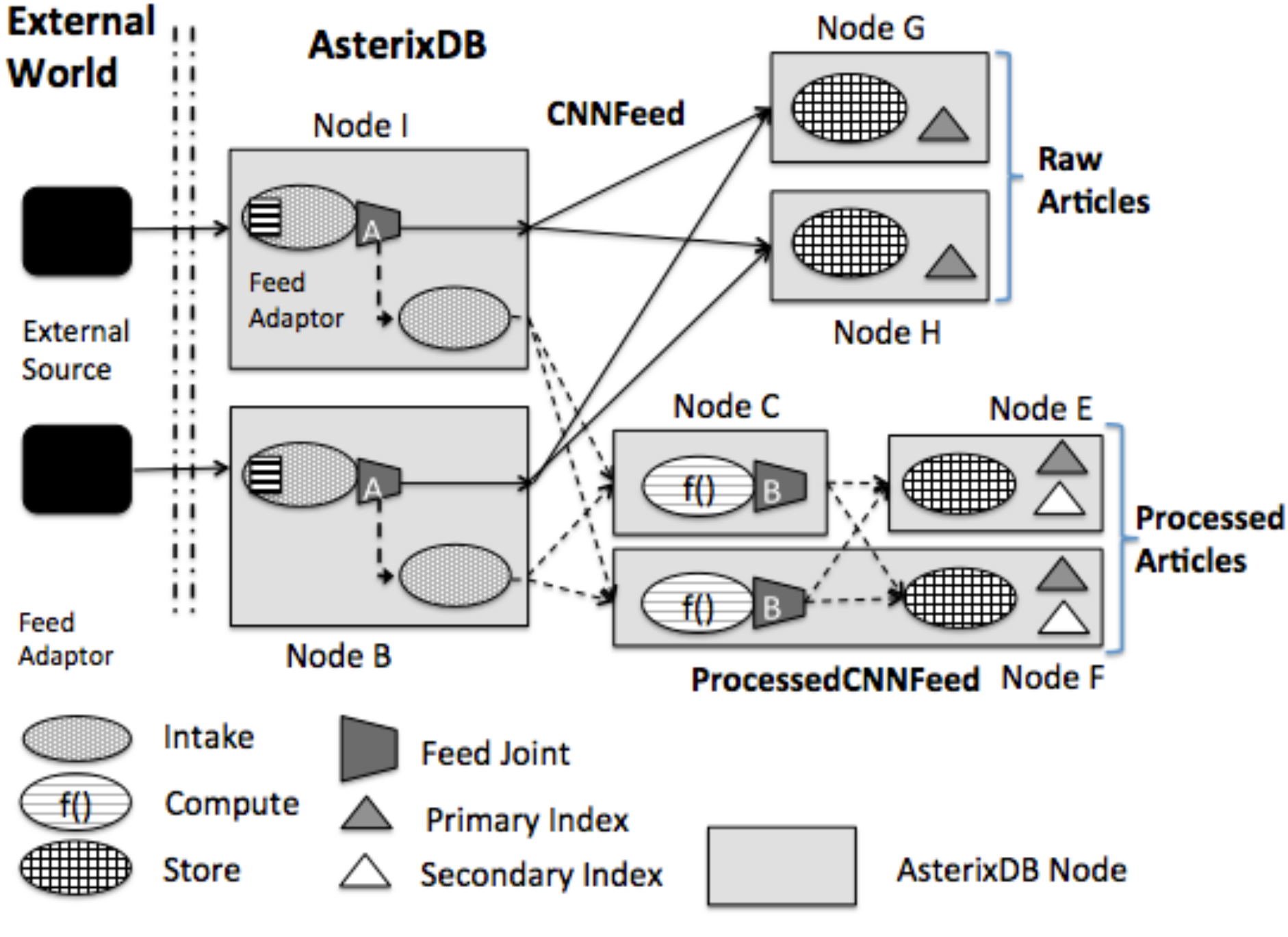}
  \label{fig:compute_node_failure_post_recovery}
  }
  
 \caption[]{Recovering from compute node failure.}
\end{figure}

\textbullet\ Compute Node Failure: We refer to Figure~\ref{fig:feed_flow_fault_tolerance} but replace the failed node A with its substitute node I. We next assume the failure of node D. Loss of node D does not impact the flow of CNNFeed but terminates the flow of data along ProcessedCNNFeed. 
To avoid data loss, the arriving records are additionally buffered (at the feed joint (kind A) in Figure~\ref{fig:compute_node_failure_intermediate_stage}) for deferred processing along the ingestion pipeline for ProcessedCNNFeed, once the pipeline has recovered. 
Since there does not exist an idle  node as before, AsterixDB must choose a substitute for node D amongst the other nodes involved in the data flow. Different options arise. Figure~\ref{fig:compute_node_failure_intermediate_stage} shows a configuration where a store node---F is chosen to act as a substitute and run a compute operator instance.  

Figure~\ref{fig:compute_node_failure_post_recovery} shows the reconfigured pipeline after data flow is restored. Alternatively, a compute node (e.g. node C) could be made to run an additional compute operator instance. The choice of the substitute node depends on dynamic parameters related to the feed pipeline such as the distribution of load across the candidate nodes, the operator that needs to be relocated and the load it is expected to add to the recipient substitute node after the movement. 

\textbullet\ Store Node Failure: Loss of a store node
translates to the loss of a partition of the dataset that is receiving the feed. AsterixDB does not yet support data replication. In absence of the replica(s), there does not exist a substitute. In the 
current implementation, a store node failure therefore results in an \emph{early} termination of an 
associated feed. As/when the failed store node re-joins the cluster and becomes available\footnote{In AsterixDB, a failed node upon re-joining the cluster undergoes a log-based recovery to ensure all hosted dataset partitions are in a consistent state.}, the feed ingestion pipeline is rescheduled. New operator instances in the rescheduled pipeline take ownership of the state left behind by their respective zombie instances from the previously failed execution. 
Data replication is on the road map of AsterixDB. An AsterixDB node hosting an in-sync replica of the lost data partition becomes the preferred choice for being an immediate substitute. The recovery phase would then involve rescheduling the pipeline to involve the replica.

%\subsubsection{Data Preservation during Ingestion}
With respect to data preservation midst hardware failures, AsterixDB does not guarantee lossless ingestion of data. Although, following a pipeline failure, operator instances save the frames from their input/output queues with the local feed manager, termination of the pipeline results in the loss of the in-flight records that failed to reach their destination. It would be possible to preserve the in-flight records by use of checkpointing to coordinate the flow of data between operators \cite{Shah:2004}. Such support can then be conditionally incorporated in a feed ingestion pipeline if the associated ingestion policy enforces lossless movement of data. 

\vspace{-0.2cm}

\section{Experimental Evaluation}
\label{sec:experimental_evaluation}
In this section, we provide an initial evaluation of the system. We study the scalability offered by the feeds support in being able to ingest an increasingly high-rate of arrival of data with the addition of resources. We also evaluate the fault-tolerance protocol by monitoring the flow of data as we introduce single and multiple hardware failures. 

\subsection{Experimental Setup}
We ran experiments on a 10-node IBM x3650 cluster. Each node had one Intel 2.26GHz processor with two cores, 8GB of RAM, and a 300GB hard disk. The following were the steps taken to prepare the experimental setup. 

(a) \emph{Modeling an External Data Source}:
We wrote a custom tweet generator, hereafter referred to as \emph{TweetGen}. TweetGen runs as a standalone process (JVM) and can be configured to output synthetic but meaningful tweets (in JSON format) at a configurable rate---\emph{tweets per second} (\emph{\textbf{twps}}). The \emph{RawTweet} datatype created in Figure~\ref{fig:feed_data_specification} showed the equivalent ADM representation for a tweet output by TweetGen. TweetGen listens for a request for data at a pre-determined port that is passed as an argument. Initiating the generation and the flow of data requires an initial handshake (by an interested receiver), subsequent to which data is ``pushed'' to the receiver at a constant rate (\emph{twps}).  

(b) \emph{Creating a feed}:
To ingest data from TweetGen, we wrote a custom socket-based adaptor---\emph{TweetGenAdaptor}. 
The adaptor is configured with the location(s) (socket address) where instance(s) of TweetGen is/are running. Each instance of TweetGen receives a request for data from a corresponding instance of \emph{TweetGenAdaptor}, thus enabling ingestion of data in parallel.

(c) \emph{Creating datasets} (\emph{and indexes}):
We used the AQL statements shown in Figure~\ref{fig:feed_dataset_definition}
 (from Section~\ref{subsec:AsterixDB_Data_Model}) to create the target datasets (and indexes) for persisting the feed. 

Figure~\ref{fig:feed_sample_experiment} shows an example setup, wherein we concurrently launch two instances of TweetGen (twps=5000) on separate machines (outside our test cluster). Next, we define a feed---\emph{TweetGenFeed}---using our custom adaptor and provide the (socket) address for each of the instances of TweetGen. Finally, we connect the feed to a dataset to trigger the flow of data. The actual experimental setup used in our experiments was similar to the example shown in Figure~\ref{fig:feed_sample_experiment}. However to evaluate the system and measure (and compare) performance parameters, we varied the size of our test cluster, the number of parallel instances of TweetGen, the twps associated with each instance and the ingestion policy. 

\vspace{-0.4cm}
\begin{figure}[!ht]
\begin{aqlschema}
10.1.0.1> java TweetGen -port 9000 -twps 5000 
10.1.0.2> java TweetGen -port 9000 -twps 5000 

create feed TweetGenFeed using TweetGenAdaptor 
("datasource"="10.1.0.1:9000, 10.1.0.2:9000"));

create secondary feed ProcessedTweetGenFeed from 
feed TweetGenFeed apply function addHashTags;

connect feed ProcessedTweetGenFeed 
to dataset ProcessedTweets;

\end{aqlschema}
\vspace{-0.4cm}
\caption{An example experimental setup}
%\vspace{-0.5cm}
\label{fig:feed_sample_experiment}
\end{figure}
 
\subsection{Scalability}
\label{sub-sec:scalability_experiment}
We first evaluated the ability of the feed ingestion support to scale and ingest an increasingly large volume of data when additional resources are added. If the record arrival rate exceeds the rate at which they can be processed and ingested in AsterixDB, the excess records thereof are either spilled to disk (for deferred processing) or discarded altogether. The precise behavior is chosen by the associated ingestion policy. For the experiment, we chose not to defer the processing of the excess tweets (by spilling them to disk) so that we may evaluate the ability to successfully ingest data as a function of available resources.  
The first statement in Figure~\ref{fig:feed_scalability_experiment} creates a custom policy that extends the built-in policy (\emph{Basic}) by overriding the policy parameter---\emph{excess.records.spill} (refer to Table~\ref{BuiltinPolicies} for a description of the policy parameter).
\vspace{-0.4cm}
\begin{figure}[!ht]
\begin{aqlschema}
create policy no_spill_policy from 
policy Basic set (("excess.records.spill","false"))

connect feed ProcessedTweetGenFeed to dataset 
ProcessedTweets using policy no_spill_policy;

\end{aqlschema}
\vspace{-0.5cm}
\caption{Creating and using a custom policy}
\label{fig:feed_scalability_experiment}
\end{figure}
\vspace{-0.2cm}

\begin{figure}[!h]
  \includegraphics[width=65mm,height=35mm]{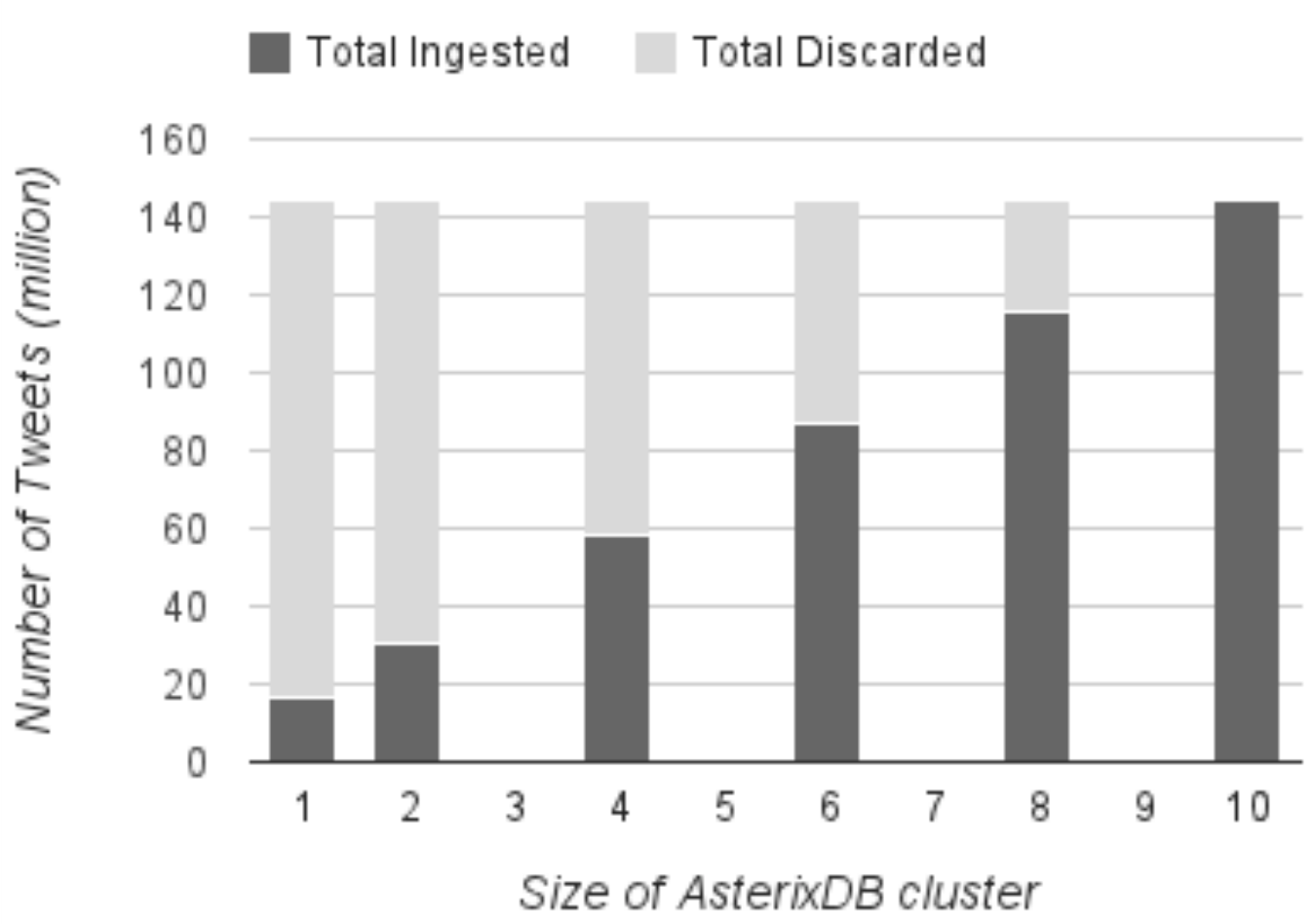}
  \caption{Scalability: Measure of the 
  number of records (tweets) successfully ingested (persisted and indexed) as the cluster size is varied}
  \label{fig:ScalabilityExperiment}
\end{figure}

In the experiment, we chose the amount of data loss as our performance metric. Ideally we don't want any data loss. So we increased the hardware until there is no data loss. A total of 6 instances of TweetGen were run on machines outside the test cluster and were configured to generate at a constant rate (20k twps) for a continuous duration of 20 minutes. We measured the total number of ingested (persisted and indexed) tweets and repeated the experiment by varying the size of our test cluster. The experimental results in Figure~\ref{fig:ScalabilityExperiment} show a significant proportion of records that are being discarded for lack of resources on a small size cluster of 1--4 nodes. On a bigger cluster, the proportion of discarded tweets declines, indicating the ability of the system to ingest an increasingly high volume of data when additional resources (nodes) are added. 
 
\vspace{-0.3cm}

\subsection{Fault Tolerance}
\begin{figure}[!h]
  \includegraphics[width=80mm,height=35mm]{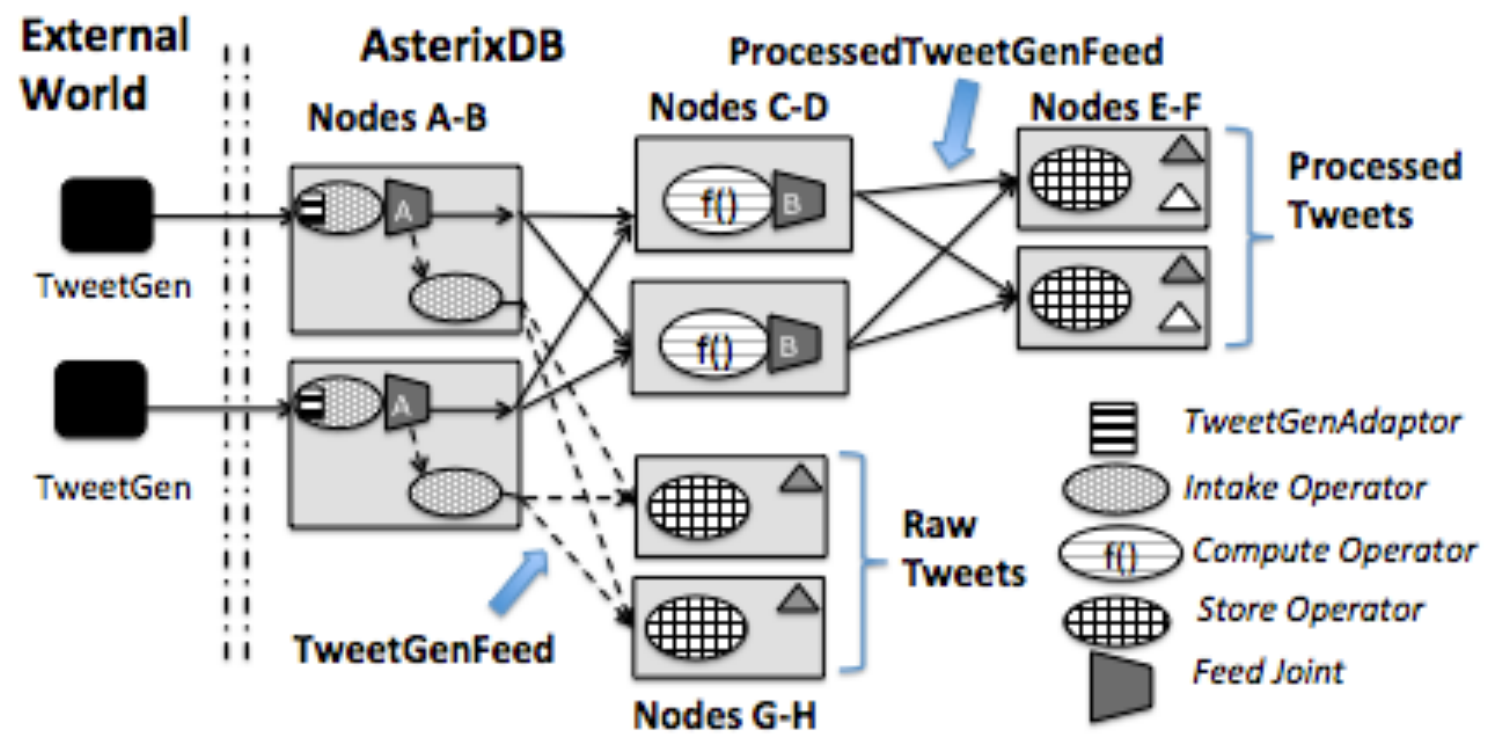}
  \caption{Feed cascade network set up for evaluation of fault tolerance protocol}
  \label{fig:FaultToleranceExperimentCascadeNetwork}
\end{figure}

We next evaluated the ability of the system to recover from single/multiple hardware failures while continuing to ingest data. This experiment involved a pair of TweetGen instances (twps=5000), each running on a separate machine and located outside the AsterixDB cluster. We connected the feeds---\emph{TweetGenFeed} and \emph{ProcessedTweetGenFeed}--to their respective target dataset and used the built-in policy---\emph{Fault-Tolerant} in doing so (Figure~\ref{fig:fault_tolerance_experiment_statements}). The nodegroup associated with each dataset included a pair of nodes.
\begin{figure}[!ht]
\vspace{-0.5cm}

\begin{aqlschema}

connect feed ProcessedTweetGenFeed to 
dataset ProcessedTweets using policy FaultTolerant;

connect feed TweetGenFeed to 
dataset RawTweets using policy FaultTolerant;

\end{aqlschema}
\vspace{-0.4cm}
\caption{Connected feeds to respective dataset}
\label{fig:fault_tolerance_experiment_statements}
\vspace{-2mm}
\end{figure}
To make things interesting and illustrate that the order of connecting related feeds is not important, we connected \emph{ProcessedTweetGenFeed} prior to connecting its parent feed---\emph{TweetGenFeed}. In absence of an available feed joint, the ingestion pipeline for \emph{ProcessedTweetGenFeed} is constructed using the feed adaptor . The physical layout of the dataflow as scheduled on our AsterixDB cluster during our experiment is shown in Figure~\ref{fig:FaultToleranceExperimentCascadeNetwork}. The ingestion pipeline for \emph{TweetGenFeed} is sourced from the feed joints (kind A) provided by \emph{ProcessedTweetGenFeed}.

\begin{figure}[!h]
 
  \subfigure[TweetGenFeed] { 	  
  \includegraphics[width=70mm,height=35mm]{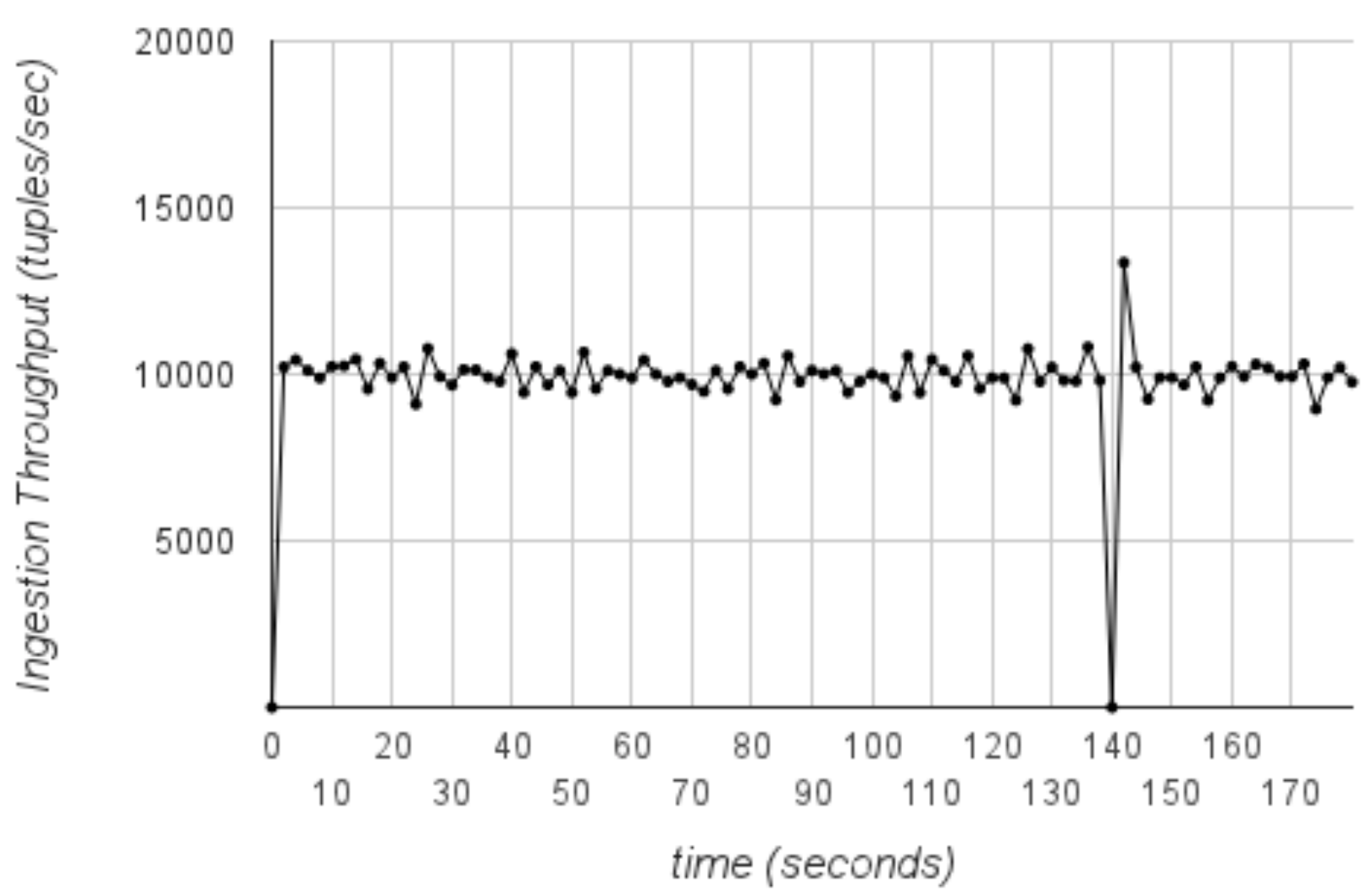}
   \label{fig:ingestion_throughput_primary_feed}  
  }

  \subfigure[ProcessedTweetGenFeed] { 	  
  \includegraphics[width=70mm,height=35mm]{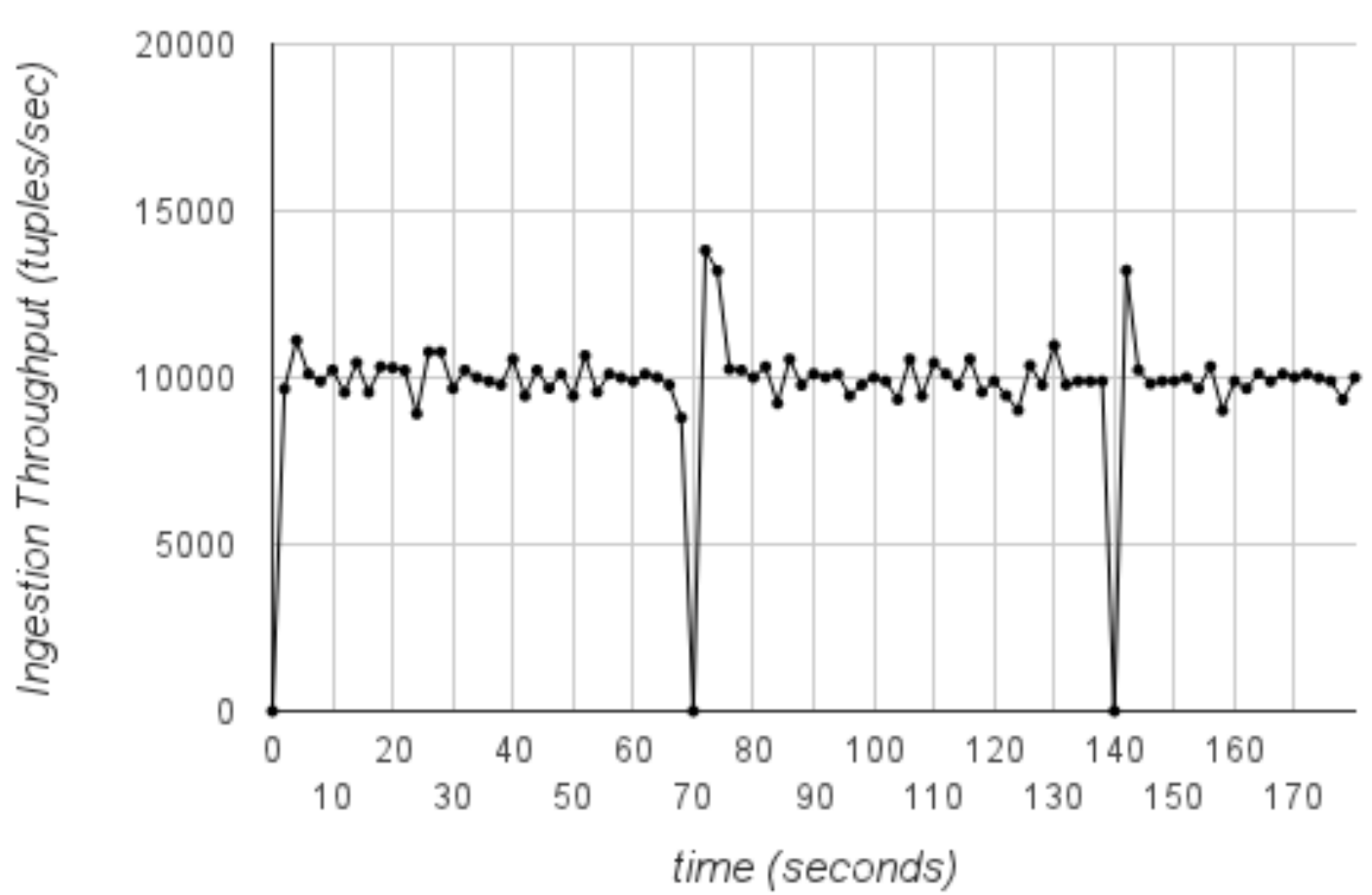}
   \label{fig:ingestion_throughput_sec_feed}  
  }
 \caption[]{Instantaneous ingestion throughput plotted on a timeline with interim hardware
 failures}
%\vspace{-0.4cm}
\end{figure}

We measured the number of records inserted into each target dataset during consecutive 2 second intervals to obtain the instantaneous ingestion throughput for the associated feed. We caused a compute node failure (node C in Figure~\ref{fig:FaultToleranceExperimentCascadeNetwork}) at t=70 seconds. This was followed by a concurrent failure of an intake node (node A) and a compute node (node D) at t=140 seconds.
The instantaneous ingestion throughput for each of the feeds---\emph{TweetGenFeed} and \emph{ProcessedTweetGenFeed}, as plotted on a timeline is shown in Figure~\ref{fig:ingestion_throughput_primary_feed} and  Figure~\ref{fig:ingestion_throughput_sec_feed} respectively. Following are the noteworthy observations.

(i) Recovery Phase: 
The failures are reflected as a corresponding drop in the instantaneous ingestion throughput at the respective times on the timeline. Each failure was followed by a recovery phase that reconstructed the ingestion pipeline and resumed the flow of data into the target dataset (within 2-4 seconds). 

%\vspace{-0.4cm}

(ii) Fault Isolation:
Data continues to arrive from the external source at the regular rate, irrespective of any failures in an AsterixDB cluster. During the recovery phase for \emph{ProcessedTweetGenFeed}, the feed joint(s) buffer the records until the pipeline is resurrected but allow the records to flow (at their regular rate) into any other ingestion pipeline that does not involve the set of failed node(s) and hence is not broken. 
This helps in ``localizing'' the impact of a pipeline failure and is a desirable feature of the system.  As shown in Figure~\ref{fig:ingestion_throughput_primary_feed}, \emph{TweetGenFeed} is not impacted by the failure of node C at t=70 seconds. Note that the buffered records are subsequently sent downstream in bulk when the failed pipeline is resurrected. This manifests as a transient positive spike in the ingestion throughput (immediately following the resumption of flow) that soon converges to the regular rate of record arrival.\\ \\

\section{Conclusion and Future Work}
\label{sec:conclusion}
We have described the native support for data feed management in AsterixDB and addressed the challenges involved in building a  fault-tolerant data ingestion facility that scales by employing partitioned parallelism. A generic plug-n-play model and provision to associate an ingestion policy helps cater to a wide variety of data sources and high-level applications. The ability to cascade feeds is useful in driving multiple applications concurrently. We provided a preliminary evaluation of the system; emphasizing the ability of the system to scale to ingest increasingly large volume of data and to handle (hardware) failures during feed ingestion. 

As an immediate next step, we plan to conduct a thorough evaluation of the system by comparing it with a `glued' together combination of systems (probably Storm and MongoDB) on grounds of complexity, performance, data consistency, fault tolerance and scalability. As part of the longer term future work, we wish to make the system `elastic' by adding the  ability to re-structure an ingestion pipeline and/or resize the cluster dynamically in response to a fluctuating workload and use of expensive UDFs. We also wish to support data replication in AsterixDB and address the associated challenges involved in replicating high-velocity data deposited by a data feed. 
\vspace{-2mm}
\section{Acknowledgements}
\label{sec:acknowledgments}
This project is supported by NSF IIS award 0910989 and NSF grant CNS-1305430. The cluster used in our experiments is supported by NSF grant CNS-1059436. R. Grover is also supported by a Yahoo! Key Scientific Challenge Award. We would like to thank John Shafer (Microsoft Research) for providing real use cases and S Sudarshan (IIT-Bombay) for his feedback on an earlier version of the paper.

\small
\bibliography{feeds_grover_carey}

%{\footnotesize\bibliography{feeds_grover_carey}}
%\bibliography{feeds_grover_carey}

\bibliographystyle{plain}
\end{document}